\documentclass[sigconf]{acmart}
\AtBeginDocument{%
  \providecommand\BibTeX{{%
    \normalfont B\kern-0.5em{\scshape i\kern-0.25em b}\kern-0.8em\TeX}}}

\copyrightyear{2022}
\acmYear{2022}
\setcopyright{acmcopyright}\acmConference[CHI '22]{CHI Conference on Human Factors in Computing Systems}{April 29-May 5, 2022}{New Orleans, LA, USA}
\acmBooktitle{CHI Conference on Human Factors in Computing Systems (CHI '22), April 29-May 5, 2022, New Orleans, LA, USA}
\acmPrice{15.00}
\acmDOI{10.1145/3491102.3501928}
\acmISBN{978-1-4503-9157-3/22/04}

\usepackage{subfigure}
\usepackage{enumitem}
\usepackage{hyperref}
\usepackage{tabularx}

\begin{document}

\title[GAN'SDA Wrap: Geographic And Network Structured DAta on surfaces that Wrap around]{GAN'SDA Wrap: Geographic And Network Structured DAta\\
on surfaces that Wrap around}


\author{Kun-Ting Chen}
\orcid{0000-0002-3217-5724}
\affiliation{%
  \institution{Monash University}
  \city{Melbourne}
  \state{VIC}
  \country{Australia}
  \postcode{3145}
}
\email{kun-ting.chen@monash.edu}

\author{Tim Dwyer}
\affiliation{%
  \institution{Monash University}
  \city{Melbourne}
  \state{VIC}
  \country{Australia}
  \postcode{3145}
}
\email{tim.Dwyer@monash.edu}

\author{Yalong Yang}
\orcid{0000-0001-9414-9911}
\affiliation{%
  \institution{Virginia Tech}
  \city{Blacksburg}
  \state{VA}
  \country{USA}
  \postcode{24060}
}
\email{yalongyang@vt.edu}

\author{Benjamin Bach}
\affiliation{%
  \institution{University of Edinburgh}
  \city{Edinburgh}
  \country{UK}
}
\email{bach@ed.ac.uk}

\author{Kim Marriott}
\affiliation{%
  \institution{Monash University}
  \city{Melbourne}
  \state{VIC}
  \country{Australia}
  \postcode{3145}
}
\email{kim.marriott@monash.edu}

\renewcommand{\shortauthors}{Kun-Ting Chen and Tim Dwyer, et al.}


\newcommand{\ben}[1]{\textcolor{red}{ben: #1}}
\newcommand{\tim}[1]{\textcolor{green}{tim: #1}}
\newcommand{\yalong}[1]{\textcolor{blue}{yalong: #1}}

\def\rev#1{#1}
\newcommand{\kt}[1]{\textcolor[RGB]{255, 165, 0}{#1}}

\newcommand{\fmap}{\textsc{Map Projection}}
\newcommand{\finteraction}{\textsc{Interaction}}
\newcommand{\fdifficulty}{\textsc{Difficulty}}

\newcommand{\xx}{\textcolor{red}{XX}}
\newcommand{\dsmall}{\textsc{Small}}
\newcommand{\dlarge}{\textsc{Large}}
\newcommand{\deasy}{\textsc{Easy}}
\newcommand{\dhard}{\textsc{Hard}}

\newcommand{\mtime}{\textsc{Time}}
\newcommand{\merror}{\textsc{Error}}
\newcommand{\mconfidence}{\textsc{Confidence}}
\newcommand{\moverall}{\textsc{Overall}}
\newcommand{\mpref}{\textsc{Rank}}



\newcommand{\linefig}[1]{
  \includegraphics[height=\fontcharht\font`\B]{#1.png}%
}

\newcommand{\dstatic}{\textsc{Static}}
\newcommand{\dinteractive}{\textsc{Interactive}}
\newcommand{\tmollweide}{\textsc{Moll\-weide Hemisphere}  \linefig{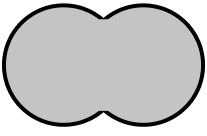}}
\newcommand{\thammer}{\textsc{Hammer} \linefig{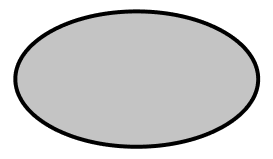}}
\newcommand{\torthographic}{\textsc{Orthographic Hemisphere} \protect\linefig{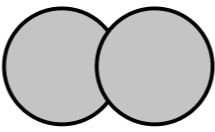}}
\protected\def\tequirectangular{\textsc{Equi\-rect\-angular} \linefig{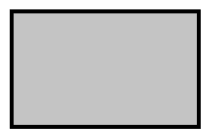}}

\newcommand{\tequalearth}{\textsc{Equal Earth} \protect\linefig{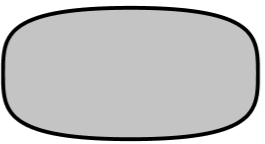}}
\newcommand{\tdistancecomparison}{\textsc{Distance Comparison} \linefig{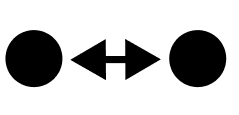}}
\newcommand{\tareacomparison}{\textsc{Area Comparison} \linefig{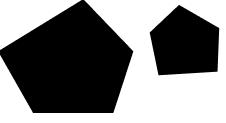}}
\newcommand{\tdirectionestimation}{\textsc{Direction estimation} \linefig{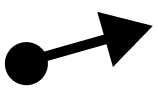}}

\newcommand{\ttorus}{\textsc{Torus} \linefig{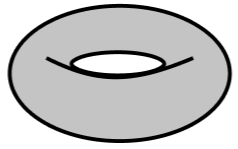}}
\newcommand{\tnodelink}{\textsc{Flat} \linefig{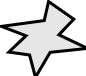}}

\newcommand{\tpairwise}{\textsc{Pairwise}}

\newcommand{\tclusteridentification}{\textsc{Cluster Number} \protect\linefig{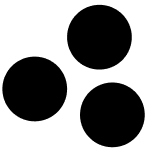}}
\newcommand{\tshortestpath}{\textsc{Shortest Path Number} \protect\linefig{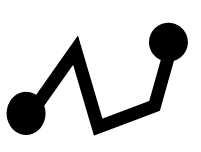}}
\newcommand{\tbelongtocluster}{\textsc{Node Cluster}}

\newcommand{\dsmalleasy}{\textsc{Small+Easy}}
\newcommand{\dlargeeasy}{\textsc{Large+Easy}}
\newcommand{\dsmallhard}{\textsc{Small+Hard}}
\newcommand{\dlargehard}{\textsc{Large+Hard}}

\begin{abstract}
There are many methods for projecting spherical maps onto the plane.  Interactive versions of these projections allow the user to centre the region of interest.  However, the effects of such interaction have not previously been evaluated.  In a study with 120 participants we find interaction provides significantly more accurate area, direction and distance estimation in such projections.
The surface of 3D sphere and torus topologies provides a continuous surface for uninterrupted network layout.  But how best to project spherical network layouts to 2D screens has not been studied, nor have such spherical network projections been compared to torus projections.  Using the most successful interactive sphere projections from our first study, we compare spherical, standard and toroidal layouts of networks for cluster and path following tasks with 96 participants, finding benefits for both spherical and toroidal layouts over standard network layouts in terms of accuracy for cluster understanding tasks.

\end{abstract}



\begin{teaserfigure}
  \includegraphics[width=\textwidth]{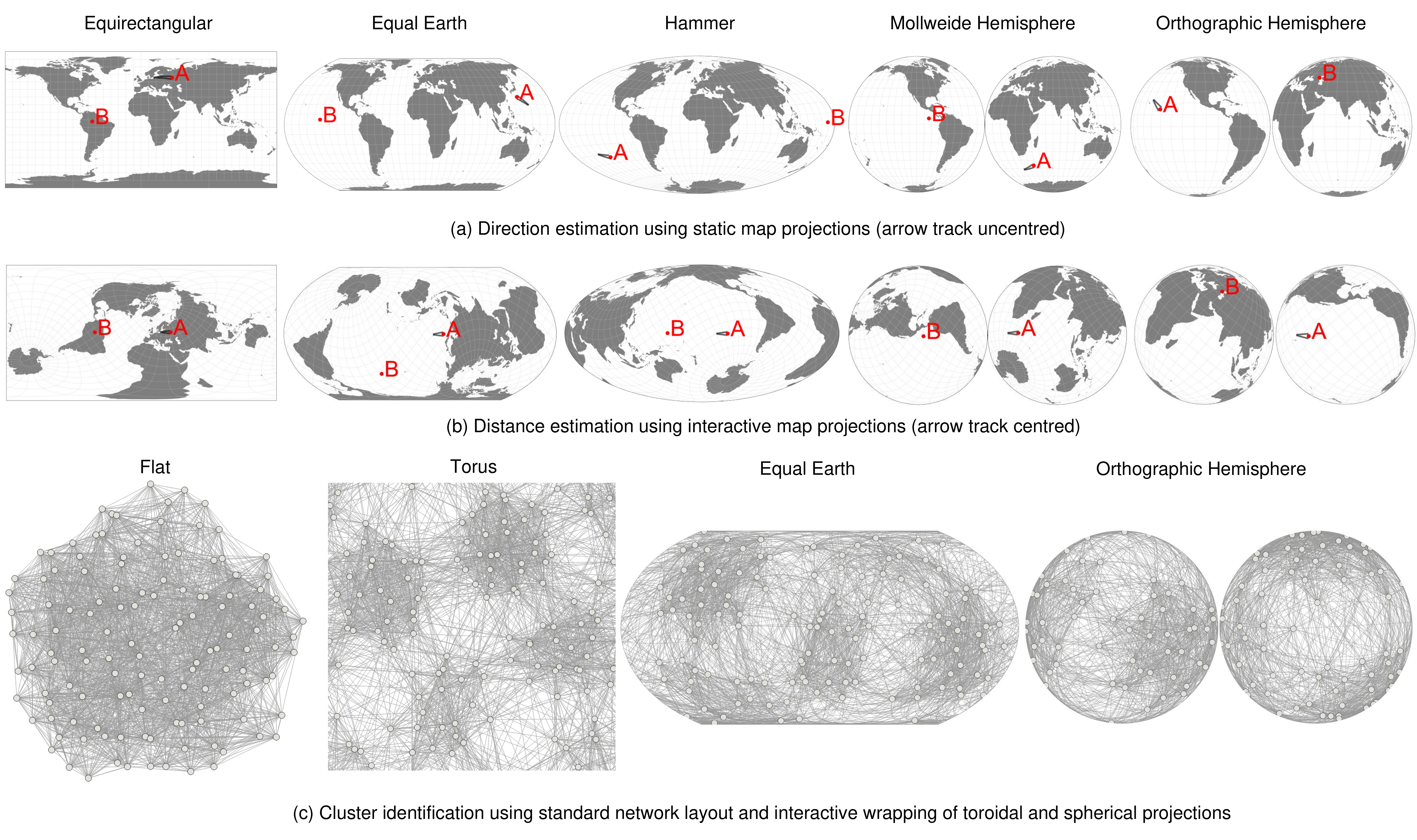}
  \caption{Top row: Direction estimation task of geographic data using the five different spherical projections evaluated in Study 1, with (a) and without (b) interaction. Bottom row: Study 2 compares the two interactive spherical projections found most effective in Study 1, and compares them to toroidal and standard `flat' layout for network data.}
  \Description{Examples of geographic and network structured data that wraps around. Top row: Direction estimation tasks of the five different spherical projections of geographic data evaluated in Study 1, with (a) and without (b) interaction. For example, for equal earth and orthographic hemisphere, the trajectory from the dot A misses dot B. Bottom row: Study 2 compares the two interactive spherical projections found most effective in Study 1, and compares them to toroidal and standard `flat' layout for network data.}
  \label{fig:teaser}
\end{teaserfigure}

\maketitle

\section{Introduction}
While most people agree that the world is spherical, most computer displays remain stubbornly flat.  To show maps of the entire earth's surface on a screen we have to somehow cut, stretch and squash it.  While cartographers and mathematicians have developed many methods to \textit{project} the surface of a sphere to a plane, distortion or discontinuity at the edges of the projection are inevitable; with different methods achieving different trade-offs between (e.g.) area, distance, or direction preservation and discontinuities~\cite{snyder1987map}.  
As a consequence distances and areas closer to the north and south poles may appear larger and the shapes more distorted than those at the centre~\cite{jenny2017guide}. Moreover, distances between  geographical locations or areas of continents or seas that are wrapped across the projection boundary are also easily misinterpreted~\cite{hennerdal2015beyond}. 

Most of these projections were originally developed for \textit{static} display of the earth in printed maps and atlases. However, with modern computer graphics we can create interactive versions of these projections, which can be panned, reoriented and recentred with simple mouse or touch drags~\cite{bostock_code_2013,Davies:2013ug}, thus allowing the viewer  to centre a region of interest so as to minimise distortion and discontinuities.

But spherical projection has application beyond geographic data.  
There can be advantages to laying out abstract data (data which has no inherent geometry) on the surface of a sphere such that there is no arbitrary edge to the display or privileged centre~\cite{rodighiero2020drawing}.  
In particular, node-link representations of highly connected network data (which is ubiquitous in the world around us, from protein-protein interaction networks, to social, communication, trade or electrical networks) have no real ``inside'' or ``outside''.  Arguably, they can be better distributed across the surface of a topology which ``wraps around'' continuously.


While the relationship between readability of geographic and network data on spherical projections has not been studied directly, there are commonalities in the analysis tasks that might be applicable for each. For example, understanding network clusters may involve comparing the relative size of their boundaries, similar to map area and shape comparison. Network path following tasks require the user to trace links (AKA `edges') between nodes in the network while maps also require understanding how regions connect, and in both case splits or distortion due to spherical projection may present a challenge.

As we discuss in detail in \autoref{sec:background}, while there has been much work to develop the algorithms for spherical projection of maps and algorithms for layout of networks on a 3D sphere, we find that interactive 2D displays of such spherical layouts of data (whether geographic or network) are not well studied. Also projections of 3D spherical layouts of network data have not been compared to projections of networks arranged on the surface of other 3D geometries, in particular toroidal layouts which have recently been studied by Chen et al.~\cite{chen2020doughnets,chen2021sa}.


Therefore, \textbf{this paper has two aims that are interlinked}:

\noindent (1) To evaluate the effect of interaction on readability of different spherical projection techniques and identify the projection techniques which best support geographic comprehension tasks, such as distance, area and direction estimation (Study 1, \autoref{sec:mapstudy}).

\noindent (2) To evaluate readability of networks laid out spherically and then projected to a flat surface using the interactive techniques found most effective in Study 1, compared to conventional flat layout, and interactive projections of toroidal layout (Study 2, \autoref{sec:networkstudy}).

\textbf{To achieve these aims many gaps in past work had to be addressed leading to seven distinct contributions:}

\noindent(1) To our knowledge, we are the first to systematically evaluate the effect of introducing interactive panning on different geographic spherical projection techniques.  We find that interaction overwhelmingly improves accuracy \rev{and subjective user preference} compared to static projections across all projection methods and tasks considered, at the cost of increased time due to interaction  (\autoref{sec:mapstudy}).

\noindent(2) For interactive projections, we find that the best of those tested depends on task, however, \torthographic{} and \tequalearth{} had  advantages in terms of speed, accuracy and qualitative feedback \rev{while \tequirectangular{} may be a poor choice even with the ability to pan} (\autoref{sec:mapstudy}).

\noindent(3) We are also the first to compare spherical network projections against toroidal and flat layouts (\autoref{sec:networkstudy}). 

\noindent(4) We adapt a pairwise gradient descent algorithm for flat and toroidal layout \cite{chen2021sa} to consider spherical distance between two nodes when laying them out on a 3D spherical surface. Cartographic projection to a 2D diagram results in 2D drawings whose links wrap around the edge of the display (\autoref{sec:layout}). 

\noindent(5) We present algorithms for computing how best to automatically rotate the spherical layout to minimise the number of links wrapped when projected (\autoref{sec:autopan}).

\noindent(6) For cluster identification tasks we find that all toroidal and spherical projections tested outperform traditional flat network layout \rev{for accuracy}, while toroidal and spherical equal earth outperform orthographic projection \rev{for accuracy, speed and subjective user rank}.

\noindent(7) For path following tasks we find that toroidal and traditional flat network layout outperform both spherical equal earth and orthographic hemisphere projections \rev{for accuracy and subjetive user rank} (\autoref{sec:networkstudy}).
    
Our results suggest that \textbf{interactive panning should be routinely provided in online maps to alleviate misconceptions arising from distortions of map projection}.  Our results also confirm the benefits of topologically closed surfaces, such as the surface of a torus or sphere, when using node-link diagrams to investigate network structure.  
This finding suggests that \textbf{interactive projections of networks arranged on 3D surfaces should be more commonly used for cluster analysis tasks} and further, that \textbf{toroidal layouts may be a good general solution}, being not only more accurate than standard flat layout for cluster tasks but also at least as accurate for path following tasks.

\rev{Collectively, we refer to the family of techniques for visualising geographic and network structured data on surfaces that wrap around by the acronym ``GAN'SDA Wrap", in a rhythmic head nod to Tupac Shakur.}
The full set of map and network study including training materials, instructions, study trials are available in the supplementary materials and associated OSF repository:~\url{https://osf.io/73p8w/}. 


\section{Background}
\label{sec:background}

Researchers have recently investigated the utility of visualisations based on projecting the surface of a torus~\cite{chen2020doughnets,chen2021sa} and a cylinder~\cite{chen2021rotate} onto a flat 2D plane. These have  shown benefits for understanding the structure of network diagrams~\cite{chen2021sa} and understanding cyclic time series data~\cite{chen2021rotate}. Crucially, these visualisations are interactive and the viewer can pan the projection so that it ``wraps'' around the plane. Providing interactive panning has previously been found to be of benefit in understanding network layouts based on torus projection~\cite{chen2020doughnets}. Here we investigate the utility of projections of the surface of a different geometric object, the sphere, on to a 2D plane. Again we allow the user to interactively pan the projection  around the plane. 

\subsection{Geographical map projections}
Projecting the surface of a sphere onto a 2D plane is not novel: cartographers have been doing this for centuries.
Cartographers and mathematicians have devised hundreds of projection techniques~\cite{snyder1997flattening}. The reason for this diversity is that  none of these techniques can be considered optimal in depicting geographic information~\cite{schottler2021visualizing}. Rather, each projection is a trade-off between preserving shape, area, distance or direction~\cite{snyder1997flattening} as it is not possible for a 2D projection to do all of these simultaneously. 

As a consequence, cartographers have invented what are called \emph{equal area projections} that preserve the relative area of regions on the globe. These include \tequalearth{}, \thammer{} and \tmollweide{} (see \autoref{fig:teaser})~\cite{vsavrivc2019equal, jenny2017guide}. They have also invented \emph{compromise projections} that do not preserve any of these criteria exactly but instead trade them off, creating a map that does not distort area, shape, distance or direction ``too much.''
These include \torthographic{} and \tequirectangular{} (see \autoref{fig:teaser}). 

Projections also differ on the shape of the projection. Some, such as \tequirectangular{} are rectangular, others, such as \tequalearth{} or \thammer{}, reduce distortion at the poles by projecting to a more ovoid shape. Some, such as the \tmollweide{} and the \torthographic{}, resemble the front and back view
of the 3D globe. Maps such as these in which the projection region is split are said to be interrupted. The \torthographic{}, in particular, has a naturalistic appearance as it shows the Earth viewed from infinity~\cite{jenny2017guide}. 

There are several user studies on the readability and user preference of map projection visualisations, e.g. ~\cite{hennerdal2015beyond, hruby2016journey, avric2015user, carbon2010earth}. 
In particular, these have found that viewers find it difficult to understand the distance or direction between two points if this requires reasoning about the discontinuity in the projection and mentally wrapping the projection around a globe to understand their relative position.  

This suggests that allowing the viewer to interactively pan the map projection so as to centre a region of interest may improve their understanding of the inherent distortion introduced by the projection and of the Earth's underlying geography. For instance, this allows them to reposition two points so that they are no longer separated by a discontinuity. Such interactive panning, also called spherical rotation~\cite{snyder1987map}, has been provided in many online maps for several years, e.g.~\cite{bostock_code_2013,Davies:2013ug,jenny_interactive_2016}. 
\rev{One recent study of pannable terrain maps found that they perform more accurately than static maps but at the cost of additional time in task completion and concluded therefore that results of existing static map reading studies are likely not transferrable to interactive maps~\cite{herman2018evaluation}.}
However, surprisingly, as far as we are aware there has been no evaluation of whether interactive panning of globe projections improves performance on standard geographical tasks such as estimating the distance or direction between two points or the relative area of two regions. 

The only direct research of pannable globes that we know of are two studies in virtual reality (VR) investigating different visualisations for understanding origin-destination flow between locations on the Earth's surface~\cite{yang_origin-destination_2019}. While it was not their main focus, the studies revealed that interactive panning improved task accuracy at the cost of time when viewing flow shown using straight lines on a flat map. 
However, it is likely that this was not because panning was used to reduce geographical distortion but rather that it was used to separate the flow lines which were the focus of the tasks. Furthermore, the static and interactive conditions were across different studies so comparison was between groups.  Here we present a more systematic and direct study of interactive panning for geographic tasks.

\subsection{Spherical representation of non-geographic data}

\rev{Researchers have also explored non-planar geometries for visualising networks, with evidence that the third dimension can be used improve readability by removing link crossings~\cite{ware2008visualizing, greffard2011visual} at the cost of known issues of three-dimensional representation (e.g.\ occlusion, readability, perspective distortion).}
Visualisation researchers have also explored the benefits of laying out node-link diagrams on to the surface of a sphere. The potential benefit is that, just as for the torus,  they are topologically closed surfaces: there is no centre or border to the surface and so it may allow the layout to better unravel the network and show its structure~\cite{brath2012sphere,perry2020drawing, rodighiero2020drawing}. Such spherical network layouts are most commonly  viewed in immersive VR (e.g.~\cite{kwon2016study}) or as perspective projection of a globe on a standard 2D monitor (e.g.~\cite{brath2012sphere,kobourov2008morphing}).
It is much less common for these to be  projected onto a 2D plane using a map projection (though some static map projections of graphs were demonstrated by~\cite{rodighiero2020drawing}). However, map projections have the great advantage over simple perspective rendering of a 3D globe, that the whole network can be seen at once (with Orthographic Hemisphere projection being the closest to a direct rendering of the 3D globe, but with most sides shown simultaneously).

To the best of our knowledge, while it is common to allow the viewer to interactively rotate a globe when shown in 3D we do not know of previous use of interactive panning of  spherical network diagrams when they have been projected onto a 2D plane using a map projection. However, we would expect similar benefits to providing interactive panning for projections based on the torus~\cite{chen2020doughnets}. 

Similar to node-link embeddings, spheres have been used to embed self-organising maps (SOMs)~\cite{wu2006spherical}, shown on both 3D representations, and projected onto a 2D-plane. Beyond node-link diagrams and SOMs, a range of other information visualisations can potentially be projected onto a sphere to minimise artefacts that occur when trying to find an optimal embedding for a 2D plane: multi-dimensional scaling (MDS) and their extension, timecurves~\cite{bach2015time}.

Rodighiero~\cite{rodighiero2020drawing} preliminarily visualised networks on a variety of spherical projections, but he did not consider spherical projection with interaction, which is essential for the users to inspect different parts of the networks.  
Manning implemented an interactive force-directed spherical layout algorithm for a variety of projections~\cite{christophermanning:force}. However, this approach used Euclidean distance between the points rather than the great circle distance on the surface of the sphere, and therefore does not take full advantage of spherical layout~\cite{perry2020drawing}.
Kobourov and Wampler give a generalisation of force-directed (based on spring-embedding) network layout to non-Euclidean topologies, including the surface of a sphere~\cite{kobourov2005non} while Perry et al.~\cite{perry2020drawing} give an algorithm based on MDS using the smacofSphere R package~\cite{de2009multidimensional}. 
Most importantly, there is no user study \rev{that} evaluates these proposed spherical graph projection techniques, and therefore there is no empirical evidence for their effectiveness.
\section{Map Projection Techniques}
\label{sec:techniques}
We chose five representative map projections for the study.
We aim to cover a wide range of distortion properties such as preservation of area, distance, shape, direction as discussed in \autoref{sec:background}, and user preference~\cite{avric2015user}. 
We demonstrate the key characteristics of the map projections in \autoref{tab:geo-techniques} and discuss the details as follows:

\begin{table*}[t]
    \centering
    \begin{tabularx}{\linewidth}{Xlllll}
    \toprule
         Projection & Type & Area-preserving & Left/right edges & Top/bottom edges & Naturalistic   \\
         \midrule
         \tequirectangular & continuous & no & straight & straight & very low  \\
         \hline
         \tequalearth & continuous & yes & curved & straight & low  \\ 
         \hline
         \thammer & continuous & yes & curved & curved & low   \\
         \hline
         \tmollweide & interrupted & yes & circular & circular & high   \\
         \hline
         \torthographic & interrupted & no & circular & circular & very high    \\
         \bottomrule
    \end{tabularx}
    \caption{Key characteristics of the five map projections tested in Study 1.}
    \Description{Key characteristics (e.g., shapes, distortion, continuity) of the five map projections tested in Study 1.}
    \label{tab:geo-techniques}
\end{table*}

\noindent\textbf{\tequirectangular{}} projects the earth onto a space-filling rectangle with the north and south poles extending along the top and bottom edges, respectively. Rectangular projection is the most widely used projection, with common variations including Mercator or Plate Carrée~\cite{jenny2017guide, avric2015user}. \tequirectangular{} projections preserve distances along all meridians and are useful when differences in latitude are measured~\cite{jenny2017guide}. However, it does not preserve the relative size of areas. Furthermore, it has been found confusing for tasks requiring understanding how the edges connect to each other, such as predicting the path of air plane routes crossing the top and bottom edge of the map~\cite{hennerdal2015beyond}. 

\noindent\textbf{\tequalearth{}} is similar to \tequirectangular{} but relaxes the rectangle with rounded corners, diminishing the strength of horizontal distortion near the poles. Furthermore, the similarly shaped Robinson projection has received good subjective ratings from viewers~\cite{avric2015user}. Unlike Robinson projection, \tequalearth{} preserves the relative size of the areas well.

\noindent \textbf{\thammer{}} further diminishes horizontal stretching by projecting the earth onto an ellipse such that the poles are points at the top and bottom. It preserves the relative size of areas and generally has similar properties to \tequalearth~\cite{yang2018maps, jenny2017guide}. The similarly shaped Mollweide projection has been shown less confusing than \tequirectangular{} when judging the continuity of air plane routes as described above~\cite{hennerdal2015beyond}. \rev{Furthermore, map readers prefer to see the poles as points rather than lines~\cite{avric2015user}.} It has also been found pleasing to many cartographers than other projections due to its elliptical shape~\cite{jenny2017guide}.

\noindent \textbf{\tmollweide{}} projects the earth onto two circles (hemispheres). This again helps diminish horizontal stretching but introduces the cost of new tears (interruption) that a viewer \rev{needs} to mentally close. 

\noindent \textbf{\torthographic{}} is also hemispheric. It has a naturalistic appearance as it shows the globe (from both sides) seen from an infinite distance. However, compared with \tmollweide{}, it is not area-preserving. 


For each projection, we created both \dstatic{} and  \dinteractive{} versions. With \dinteractive{}, a user can freely move regions of interest to the centre of the projection, thus reducing their distortion. Examples of the effect of interaction are shown in the top row in \autoref{fig:teaser}.

We used D3 libraries for creating all the map projection techniques. For \dinteractive{}, we follow Yang et al.~\cite{yang2018maps} and use versor dragging which controls three Euler angles. This allows the geographic start point of the gesture to follow the mouse cursor~\cite{Davies:2013ug}.

We implemented \torthographic{} with two Orthographic map projections with one showing the western and the other showing the eastern hemisphere. They are placed close together as shown in the top row of \autoref{fig:teaser}. When one hemisphere is dragged, the Orthographic projection of the other hemisphere automatically adjusts three-axis rotation angles such that it shows the correct opposite hemisphere. 

Spherical rotation of all projections is demonstrated in the supplementary material video and the OSF repository\footnote{Interactive examples can be found in \url{https://observablehq.com/@kun-ting}.}

\section{Study 1: map projection readability}
\label{sec:mapstudy}

The goal of our first study was to understand the effectiveness of (i) different projections for geographic data as well as the (ii) benefit of interactively changing the centre point of these projections. 

\subsection{Techniques}
The techniques are \tequirectangular, \tequalearth, \thammer, \tmollweide, and \torthographic, described in \autoref{sec:techniques}. Each technique is given both static images without the ability to rotate, and with interactive spherical rotation, using the mouse. A user can rotate the visualisation such that when the view is panned off one side of the display, it either reappears on the opposite side (left-right) or is horizontally mirrored (top-top, bottom-bottom), as shown in the top-row of \autoref{fig:teaser}. The area of the rectangular bounding box of each map projection condition was \rev{fixed} at 700 $\times$ 350 pixels.

\subsection{Tasks \& Datasets}
We selected three representative \rev{geographic data visualisation tasks}. They were also used in existing map projection studies~\cite{yang2018maps,hennerdal2015beyond,hruby2016journey,carbon2010earth}. 

\noindent\textbf{\tdistancecomparison:} \textit{Which pair of points (pair \textit{A} or pair \textit{B}) represents the greater geographical distance on the surface of a globe?} Participants had to compare the true geographical distance (\textit{as the crow flies}) between two pairs of points on the projection. Participants were provided with radio buttons to answer \textit{A}, \textit{B}, or \textit{not sure}. We created data sets for two difficulties through extensive piloting, based on the difference of point distance across point pairs: 10\% difference (easy) and 5\% difference (difficult). \rev{The geographic point pairs were randomly chosen, constrained by their individual angular distance (measured in differences in geographical coordinates) between 40\textdegree{} (approx. 4444\textit{km}) and 60\textdegree{} (approx. 6666\textit{km})~\cite{yang2018maps}, and a minimum 60\textdegree{} distance across point pairs. We did not set any upper-bound to not to bias any projection.} We created an additional quality control trial with 40\% difference of node distance to test a participant's attention. \rev{An example is shown in \autoref{fig:Taskscomparison}(a,c).}

\noindent\textbf{\tareacomparison:} \textit{Which polygon (\textit{A} or \textit{B}) covers the greater geographical area on the surface of a globe?} Participants had to compare the size (area) of the polygons. Participants were provided with radio buttons to answer \textit{A}, \textit{B}, or \textit{not sure}. Again, we created data sets for two difficulties based on the difference in area they cover: 10\% difference (easy) and  7\% difference (difficult). \rev{Eight geographic points of convex polygons were randomly chosen using the same method as Yang et al.~\cite{yang2018maps}, constrained by the individual geographic area between 40 and 60, and a minimum 60\textdegree{} angular distance between centroids of pairwise polygons. There is no upper-bound for the same reason as above.} We create an additional quality control trial with 40\% difference in area to test a participant's attention. \rev{An example is shown in \autoref{fig:Taskscomparison}(b,d).}

\noindent \textbf{\tdirectionestimation:} \textit{Does the trajectory of dot A hit or miss dot B on the surface of a globe?} Participants had to assess whether the trajectory, indicated by an arrow track, from point \textit{A} hits or misses point \textit{B}. Participants were provided with radio buttons to answer \textit{Hit}, \textit{Miss}, or \textit{not sure}. \rev{We randomly created pairs of geographic points (A, B, and arrow head) with a minimum angular distance of 60\textdegree{} between A and B. There is no upper-bound. For trials where the trajectory of A misses B, the angular distance between trajectory and B was constrained to 40\textdegree{}.} Examples of this task is shown in the first two rows of \autoref{fig:teaser} for each projection techniques, \rev{where A misses B for equal earth and orthographic hemisphere}. For this task, there was only one level of difficulty. We create an additional quality control trial with dot A at the centre and arrow track hitting to a dot B aligned horizontally at the centre to test a participant's attention. 

\subsection{Hypotheses}
Our hypotheses were pre-registered with the Open Science Foundation:~\url{https://osf.io/vctfu}. 

\noindent\textbf{H1-1}: \textit{\dinteractive{} has \rev{a lower error rate} than all static map projections across all tasks.} Our intuition is that interaction allows regions of interest to be centred and, thus, their distortion reduced.

\noindent \textbf{H1-2}: \textit{\dinteractive{} projections have longer task-completion time than static across all projections and tasks.} Users will spend time interacting to find an optimal centre point for each projection to solve the task.

\noindent \textbf{H1-3}: \textit{For interactivity, users prefer \dinteractive{} to \dstatic{} across all tasks.} Intuition as above.

\noindent \textbf{H1-4}: \textit{\dinteractive{} \torthographic{} has \rev{a lower error rate} than other interactive non-hemisphere (\tequirectangular{}, \tequalearth{}, \thammer{}) projections across all tasks.}
This is inspired by a virtual globe study by Yang et al.~\cite{yang2018maps}. Our intuition is that an interactive view of the 3D globe will have similar benefits to the VR representation.

\noindent \textbf{H1-5}: \textit{Users prefer \dinteractive{} \torthographic{} over all other \dinteractive{} projections across tasks.} Intuition as above.





\subsection{Experimental Setup}

Design is within-subject per task, where each participant performed \textit{one} task (\tdistancecomparison, \tareacomparison, \tdirectionestimation) on all projections (\tequirectangular, \tequalearth, \thammer, \tmollweide, \torthographic) in both \dstatic{} and \dinteractive{} and in all levels of difficulty (\textit{easy}, \textit{hard}). Each of these 10 conditions for \tareacomparison{} and \tdistancecomparison{} was tested in 12 trials \rev{with two difficulty levels} (6 easy, 6 hard). \rev{For \tdirectionestimation{}, we reduced the number of trials to 8 as pilot participants reported the direction estimation was too difficult for long distance.} 
Similar to existing visualisation crowd-sourced studies~\cite{chen2021rotate, brehmer2018visualizing}, we randomly inserted a quality control trial with low difficulty in addition to normal trials to each condition to test participants' attention. The study was blocked by interactivity. Within each block, the order of the map projection techniques  was counterbalanced using William et al.'s Latin-square design~\cite{williams1949experimental}. This technique resulted in 10 possible orderings for the 5 projections while the order of projections in each block was the same. Each recorded trial had a timeout of 20sec, inspired from pilot studies.  

\subsection{Participants and Procedures}

We crowd-sourced the study via the \textit{Prolific Academic system}~\cite{palan2018prolific}. Participants on Prolific Academic have been reported to produce data quality comparable to Amazon Mechanical Turk~\cite{peer2017beyond}. Many visualisation studies have used this platform in the past~\cite{satriadi2021quantitative, chen2021rotate}. 
We hosted the online study on Red Hat Enterprise Linux (RHEL7) system. We set a pre-screening criterion on performance that required a minimum approval rate of 95\% and a minimum number of previous successful submissions of 10. We also limited our study to desktop users with larger screens. We paid \pounds5 (i.e.\ \pounds7.5/h), considered to be a good payment according to the Prolific Academic guidelines.

We recorded 120 participants who passed the attention check trials, completed the training and recorded trials. This comprised 4 full counterbalanced blocks of participants (10 $\times$ 4 $\times$ 3 tasks). 57 of our participants were females, 63 were males. The age of participants was between 18 and 55. 7 participants rank themselves as \textit{regularly} using GIS or other tools to analyse geographical data. 105 occasionally read maps, e.g., using Google Map or GPS systems. 8 had very little experience with any sort of maps. 

Before starting the study, each participant had to complete a tutorial explaining projection techniques and tasks. 
The tutorial material contained \textit{Tissot's indicatrix}~\cite{snyder1987map}, a set of circular areas placed on both the poles the equator, indicating the type and magnitude of \rev{area, shape, and angular} distortion in a given projection. The setting has been inspired by Yang et al.~\cite{yang2018maps}. Specific instructions were given for each task, available in our supplementary material\footnote{An online demonstration of the study is available: \url{https://observablehq.com/@kun-ting/gansdawrap}}.

\subsection{Dependent Variables}

We measured \textbf{task-completion-\emph{time}} (\mtime{}) for each trial in milliseconds, counted between the first rendering of the visualisation and the mouse click of the \textit{answer trial} button, which hid the visualisation and showed an interface for the participants to input their answers. We measured the \textbf{\emph{error rate}} (\merror{}) as the ratio of incorrect over all answers.
We asked participants to \textbf{\emph{rank}} (\mpref{}) each map projection individually \rev{within the \textit{static} and the \textit{interactive} block according to their perceived effectiveness.} We also ask participants to provide their justifications for the rankings as \textbf{\emph{qualitative feedback}.} \rev{After they completed both blocks, we recorded participants' preference of the interactivity between \dstatic{} and \dinteractive{} individually for each projection, and their overall preference between \dstatic{} and \dinteractive{}.}

\subsection{Statistical Analysis Methods}
\label{sec:study-1-stats}

\begin{figure*}
    \centering
    \includegraphics[width=\linewidth]{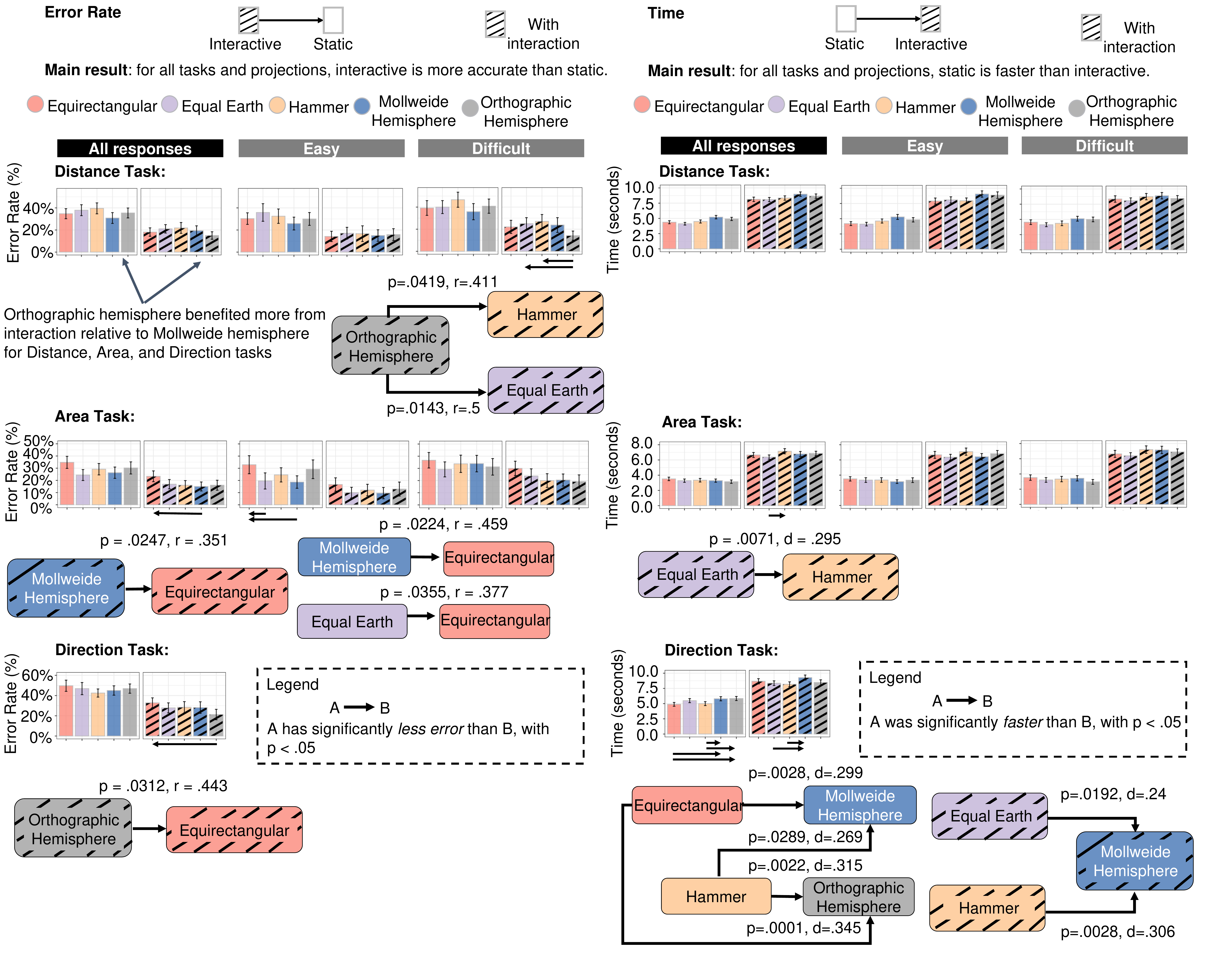}
    \Description{This diagram summarises the results of study 1 for error rate (left) and time (right). Significant differences between projections are shown as arrows. Significant differences between interactive and non-interactive conditions are omitted to improve readability. Error bars indicate 95\% confidence
intervals. Overall, equal earth and orthographic hemisphere performed well. The detailed statistical results and effect sizes can be found in Section A.1.}
    \caption{Error rate (left) and Time (right) results. Significant differences between projections are shown as arrows. Significant differences between interactive and non-interactive conditions are \textbf{omitted} to improve readability. \rev{Error bars indicate 95\% confidence
intervals. Effect size results for Cohen's r and Cohen's d~\cite{cohen2013statistical} are presented for \merror{} and \mtime{}, respectively. Statistically significant results are highlighted in flow diagrams below the bar charts. Bars and boxes with stripe pattern refer to interactive conditions. Overall, equal earth performed well for \mtime{} and \merror{} for some tasks. Orthographic hemisphere benefited more from interaction than mollweide hemisphere for \merror{}.}}
    \label{fig:study-1-results}
\end{figure*}

\begin{figure}
    \centering
    \includegraphics[width=\linewidth]{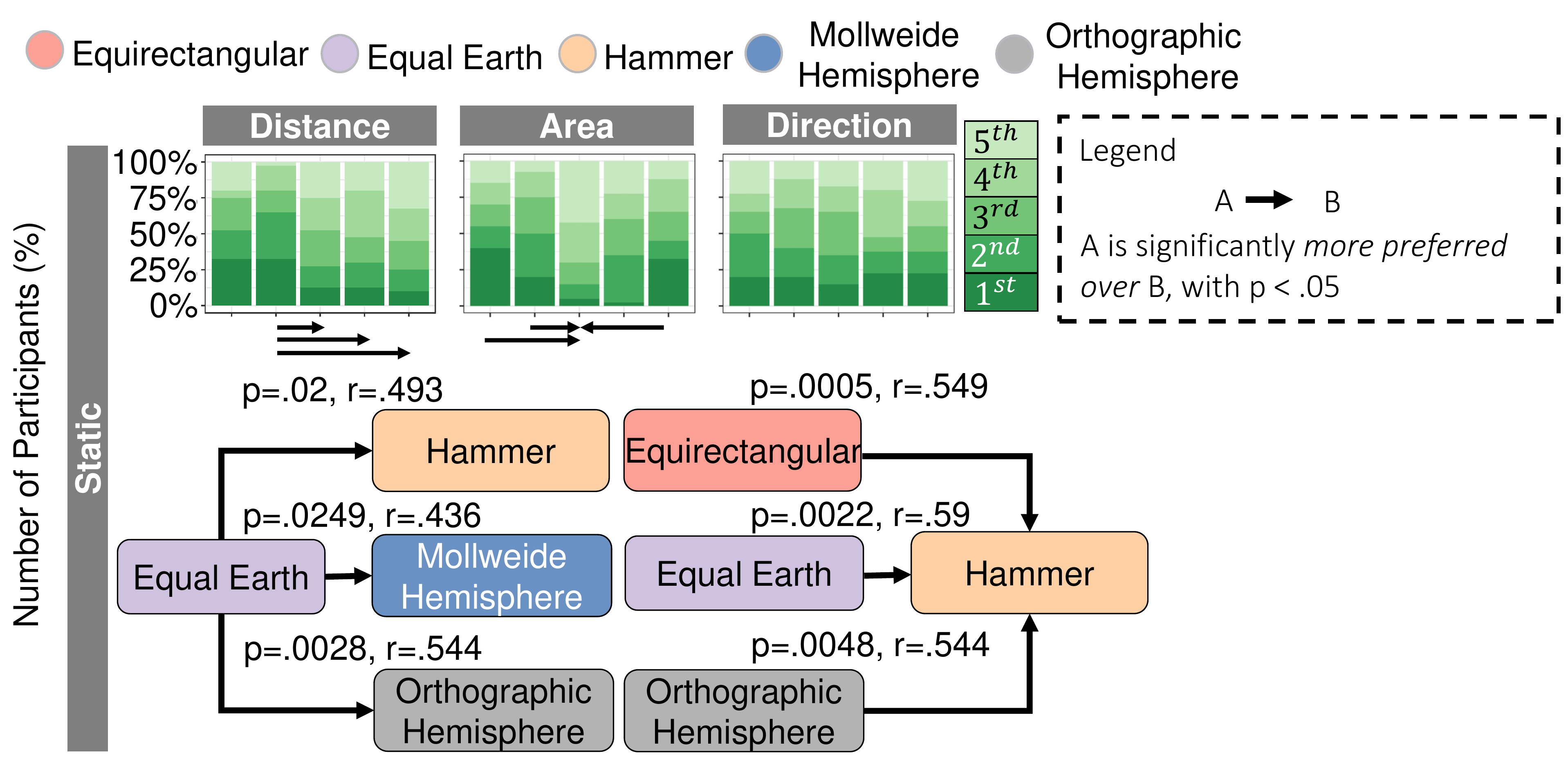}
    \Description[Legend]{This diagram shows the results of subjective user rank of map projections within the static group for three tested tasks. Lower rank indicates stronger preference. Arrows indicate statistical significance with $p<0.05$. Overall, equal earth was found to be preferred over hemispheric projections for the distance comparison task, while hammer was preferred over by other projections for area comparison task.}
    \caption{Subjective user rank of map projections within the \emph{static} group for three tested tasks. Lower rank indicates stronger preference. Arrows indicate statistical significance with $p<0.05$. \rev{Overall, equal earth was found to be preferred over hemispheric projections for the distance comparison task, while hammer was preferred over by other projections for area comparison task.}}
    \label{fig:study-1-ranking-results}
\end{figure}

\begin{figure*}
    \centering
    \includegraphics[width=\linewidth]{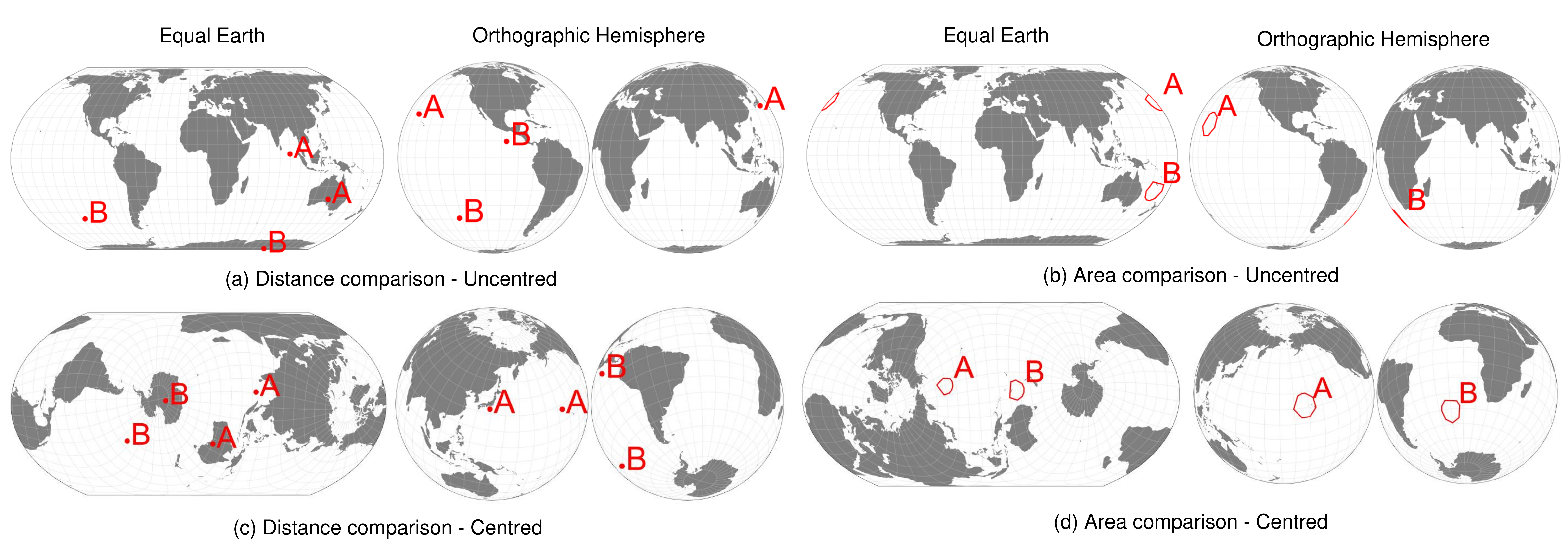}
    \Description[]{This diagram shows examples of study trials of Equal Earth and Orthographic Hemisphere for distance and area comparison tasks. \dstatic{} (a,b) may result in uncentred view. \dinteractive{} (c,d) allows a user to drag the map to find the best angle (centred) to answer the task.}
    \caption[]{Sample trials of Equal Earth and Orthographic Hemisphere for distance and area comparison tasks. \dstatic{} (a,b) may result in uncentred view. \dinteractive{} (c,d) allows a user to drag the map to find the best angle (centred) to answer the task. Direction tasks are shown in \autoref{fig:teaser}.}
    \label{fig:Taskscomparison}
\end{figure*}

We used \emph{sqrt}-transformation for \mtime{} to meet the normality assumption. We then used linear mixed modelling to evaluate the effect of independent variables on the dependent variables~\cite{Bates2015}.
We modelled all independent variables (five map projections, two interaction levels and two difficulty levels) and their interactions as fixed effects. 
We evaluated the significance of the inclusion of an independent variable or interaction terms using log-likelihood ratio. 
We then performed Tukey's HSD post-hoc tests for pairwise comparisons using the least square means~\cite{Lenth2016}. 
We used predicted vs. residual and Q---Q plots to graphically evaluate the homoscedasticity and normality of the Pearson residuals respectively.
For \merror{} and \mpref{}, as they did not meet the normality assumption, we used the \emph{Friedman} test to evaluate the effect of the independent variable, as well as a Wilcoxon-Nemenyi-McDonald-Thompson test for pairwise comparisons.
We also used Wilcoxon signed-rank test for comparing the accuracy of static and interactive map projections. \rev{The confidence intervals are 95\% for all the statistical testing.} We demonstrate the \emph{error rate} and \emph{time} in \autoref{fig:study-1-results}. We show \rev{interactivity and map projection} rankings as stacked bar charts in \autoref{fig:study-1-ranking-results}, Section A.1 (Appendix): Figure 9, and Figure 10.

\rev{Following existing work which reports statistical results with standardised effect sizes~\cite{okoe2018node,TransparentStatsJun2019, yoghourdjian2020scalability} or simple effect sizes with confidence intervals~\cite{besancon:hal-01436206, besanccon2017pressure}, we interpret the standardised effect size for a parametric test using Cohen's d classification, which is 0.2, 0.5, and 0.8 or greater for small, moderate, and large effects, respectively~\cite{cohen2013statistical}. For non-parametric test, we interpret the standardised effect size for a Wilcoxon's signed-rank test using Cohen's r classification, which is 0.1, 0.3, and 0.5 or greater for small, moderate, and large effects, respectively~\cite{cohen2013statistical, pallant2013spss}.}

\subsection{Key Findings and Discussion}
\label{sec:mapstudydiscussion}

We report on the most significant findings for \tdistancecomparison{}, \tareacomparison{}, and \tdirectionestimation{} visually in \autoref{fig:study-1-results} and \autoref{fig:study-1-ranking-results}. \rev{Detailed statistic results and pairwise effect sizes are provided in the appendix: Section A.1 in the supplementary materials}.

\textbf{Interaction improved error rate \rev{and is preferred} over static map projections while taking participants longer time to complete (across all tasks).} 
This \rev{main effect} was found statistically significant \rev{with moderate and large effect sizes} in all three tested tasks (\autoref{fig:study-1-results}, Section A.1: Figure 10, Figure 11-13). \rev{For \tdistancecomparison{} and \tdirectionestimation{}, \dinteractive{} is always better (significantly less error and more preferred) than \dstatic{}. For \tareacomparison{}, \dinteractive{} has less error than the corresponding non-interactive projection but not necessarily all other non-interactive projections. \dinteractive{} was significantly preferred (moderate effects) for \thammer{} and \tequalearth{} but not necessarily other projections for \tareacomparison{}. However, \dinteractive{} was significantly preferred in the overall user rank over \dstatic{} (top horizontal bar in Section A.1: Figure 10) for \tdistancecomparison{} (large effects), \tdirectionestimation{} (large effects) and \tareacomparison{} (moderate effects).} \dinteractive{} is always slower (large effects) than \dstatic{} (\autoref{fig:study-1-results}-Time). Therefore, we accept H1-1, H1-2, H1-3. 

These results provide strong evidence that, with interactions, participants were able to find a better projection centre than the default one in a static map. \rev{Some participants explicitly mentioned the benefits of having interaction and their preference, e.g., \emph{``when moved [,the interactive conditions makes] it easier to judge when [the areas] were both placed in the middle in the least distorted part of the map.''} (P20, Area-\dinteractive{}). and \emph{``The fact I couldn't move the pictures was frustrating [for the static conditions] and I think I didn't get many of the guesses right.''} (P4, Direction-\dstatic{}).} 




\rev{\textbf{The choice of projections makes less difference and depends on tasks. However, overall, we found that equal earth and orthographi hemisphere performed well, while equirectangular may be a poor choice, organised in the following key findings.}}


\textbf{In continuous projections, \tequalearth{} \rev{performed well in terms of \merror{} for \tareacomparison{}, \mtime{} for \tareacomparison{} and \tdirectionestimation{}, and \mpref{}-\dstatic{} for both \tdistancecomparison{} and \tareacomparison{}. \tequalearth{} was not significantly worse than any other continuous projections.}} \rev{We found that for static projections, \tequalearth{} tended to be more accurate (moderate effects) than \tequirectangular{} for Area-Easy. With interaction, \tequalearth{} tended to be faster (small effects) than \tmollweide{} for \tdirectionestimation{}, and \thammer{} for Area-All (\autoref{fig:study-1-results}-Time). Though the time results are statistically significant, the small effect sizes seem to indicate that the choice of projection makes a slight difference~\cite{helske2021can,cockburn2020threats,TransparentStatsJun2019, schafer2019meaningfulness}. For \mpref{}-\dstatic{}, \rev{\tequalearth{} was significantly preferred over \thammer{} (moderate effect), \tmollweide{}\break (moderate effect), and \torthographic{} (large effect) for \tdistancecomparison{} (\autoref{fig:study-1-ranking-results}-Distance). Furthermore, there is a strong evidence with important effects for \mpref{}-\dstatic{} that \tequirectangular{}, \tequalearth{}, and \torthographic{} were significantly preferred over \thammer{} (large effects) for \tareacomparison{} (\autoref{fig:study-1-ranking-results}-Area).} With interaction, there is no significant differences in user preference. This result is omitted from the paper and is available in Section A.1: Figure 9.} 

User preference of \dstatic{} \tequalearth{} partially confirms Šavrič et al.~\cite{avric2015user}, who found \rev{Robinson projection (which is similarl to \tequalearth) was preferred over interrupted projections such as \tmollweide{} and Goode Homolosine. This is also supported by our participants' feedback where continuous maps are preferred over interrupted ones for \dstatic{} (Section A.1.1).} 

Surprisingly, although \thammer{} is an equal-area projection, participants did not like it for area comparison in static maps (\autoref{fig:study-1-ranking-results}-Area), \rev{This result partially differs from existing studies~\cite{avric2015user} where poles represented as points were preferred over poles represented as lines}. 
We conjecture this effect is because when the target area is at the edges, the shapes are severely distorted, which makes it difficult to accurately accumulate its represented area, as participants mentioned \rev{\emph{``Equirectangular only was distorted from top to bottom, while [hammer was] also distorted on the sides.''} (P31, Area-\dstatic{}). More quotes can be found in Section A.1.1.} 

\rev{\textbf{In hemispheric projections, interaction reduced \merror{} of \torthographic{} to a point that it tended to have a lower error rate than some interactive continuous projections for Distance-Hard and Direction, and not significantly slower than any interactive projections across all tasks.} We found \torthographic{} performed well. \torthographic{} benefited more from interaction (large effects) for \merror{} than \tmollweide{} (moderate effects) for \tdistancecomparison{}, while \torthographic{} benefited slightly more from interaction with similar effect sizes for \merror{} than \tmollweide{} for \tareacomparison{} and \tdirectionestimation{} (\autoref{fig:study-1-results}-\merror{}, Section A.1: Figure 11-Figure 13). This is also supported by a strong evidence that \dinteractive{} \torthographic{} has a lower error rate (large effects) than \tequalearth{} and a lower error rate (moderate effects) than \thammer{} for Distance-Hard (\autoref{fig:study-1-results}-Error). By contrast, even with interaction, \tmollweide{} was still slower (small effects) than non-hemisphere for \tdirectionestimation{} (\autoref{fig:study-1-results}-Time). \dinteractive{} \torthographic{} did not have a significantly lower error rate for \tareacomparison{} than any interactive continuous projections, nor was there any significant difference in \mpref{}-\dinteractive{}  (Figure 9). Therefore, we reject H1-4, H1-5.} 

It was surprising that \torthographic{} \rev{was comparable to other interactive projections} for \tareacomparison{} since it is not an equal-area map projection. Meanwhile, the other non-equal-area map projection, i.e., \tequirectangular{} tended to produce more errors (moderate effects). We believe \torthographic{} were perceived as less distorted than the other projections due to the ``natural'' orthographic distortion, which is similar to viewing the sphere at infinite distance (e.g.\ as if through a telescope)\rev{, echoed by our participants (Section A.1.1)}. 

\rev{Despite being hemispheric, \tmollweide{} was found to be less intuitive for some tasks by participants (Section A.1.1). We conjecture that there is a slight distortion near the edge of two circles which may make it confusing when centring the region of interest.} However, there were also participants who did not like the hemispheric projections \rev{due to the need to inspect two separated spheres and instead they preferred the continuous maps in the non-hemispheric group (Section A.1.1).} 



\rev{\textbf{Even with interaction, \tequirectangular{} still tended to perform poorly in terms of \merror{} for \tareacomparison{} and \tdirectionestimation{}}. Overall, for static projections, \tequirectangular{} tended to have a higher error rate (moderate effects) than \tequalearth{} and \tmollweide{} (\autoref{fig:study-1-results}-Error).} 
\rev{To our surprise, with the ability to rotate to centre the region of interest, \tequirectangular{} still tended to be outperformed for \merror{} by \tmollweide{} (moderate effect) for Area-All and by \torthographic{} (moderate effect) for \tdirectionestimation{} (\autoref{fig:study-1-results}-Error)}. This partially confirms Hennerdal et al.'s static map study where \dstatic{} \tequirectangular{} was found confusing when estimating the airplane route that wraps around~\cite{hennerdal2015beyond}. We conjecture that \tequirectangular{} features the most distortion of all tested projections due to the high level of stretching near the poles, supported by participants' feedback (Section A.1.1).

\subsection{Limitations}
\label{sec:study-1-threats}


\rev{The statistically significant results with large effect sizes provide a strong evidence that adding the spherical rotation interaction to static maps improves the accuracy, is strongly preferred, but at the cost of longer completion time across all tasks. However, despite being statistically significant, the differences between projections within static or interactive groups are of small sizes for \mtime{}, medium-sized for \merror{} and large-sized for \mpref{}-Static-Area (\autoref{fig:study-1-results}, \autoref{fig:study-1-ranking-results}). Although medium-sized differences in \merror{} are likely to be noticeable in practical applications, these results only allow us to say \tequalearth{} and \torthographic{} performed well for some tasks~\cite{helske2021can,cockburn2020threats,TransparentStatsJun2019, schafer2019meaningfulness}.} 

\rev{Although the results show statistically significant differences between the selected hemispheric and non-hemispheric projections across all tasks for \merror{}, the results do not allow us to say that hemispheric projections are always more accurate than non-hemispheric projections. Arguably, the poor performance of non-hemispheric projections for \tareacomparison{} and \tdirectionestimation{} is entirely based on \tequirectangular{} compared with either \tmollweide{} or \torthographic{}. If the results from the \tequirectangular{} were not considered, the hemispheric and non-hemispheric projections seem to have very similar error rates for \tareacomparison{} and \tdirectionestimation{}. Similarly, the results do not allow us to say that non-hemispheric projections are always faster, as this seems only based on \tmollweide{} being significantly slower (small effects) than some non-hemispheric projections for \dstatic{} and \dinteractive{} for the direction tasks.}

Surprisingly, unlike the study by Yang et al.~\cite{yang2018maps}, we did not identify superior performance of \torthographic{} compared to other map projections.
We conjectured that rendering \torthographic{} on a 2D flat display produced different effectiveness in perception and interaction compared to Yang et al's 3D globes in VR.
Meanwhile, although not in all tasks, \torthographic{} demonstrated some advantages \rev{for \tdistancecomparison{}, \tdirectionestimation{} and benefited more from interaction than \tmollweide{} for \merror{} (~\autoref{sec:mapstudydiscussion})}.

\section{Spherical Network Layout}
\label{sec:layout}
In Study 1, we found that interaction (panning by spherical rotation) makes spherical geographic projections overwhelmingly more accurate for distance, area and direction tasks.  A question\rev{, however,} is whether such 2D interactive spherical projections are also useful for abstract data.  As discussed in \autoref{sec:background}, there has been a number of systems using immersive environments to visualise network data on 3D spherical surfaces or straightforward perspective projections of spheres.  \rev{Various advantages have been claimed to the opportunities for embedding a network layout in the surface of a sphere---without boundary---such as centring any node of interest in the layout~\cite{brath2012sphere,perry2020drawing, rodighiero2020drawing}, improving readability by reducing link crossings using the third dimension~\cite{ware2008visualizing}, and stereoscopy outperforms standard 2D layouts for highly overlapping clustered graphs~\cite{greffard2011visual}.}  However, the visualisation design literature cautions against such use of 3D if there are layout approaches better suited to the plane \cite[Ch.~6]{munzner2014visualization}.  

Further, there are obvious disadvantages to projection, since all projections introduce some degree of distortion and discontinuity.  There are therefore three questions: 

\noindent \textbf{RQ1:} \textit{Which of the most promising projections from our first study are \rev{the best for visualising the layout of a node-link diagram}?} 


\noindent \textbf{RQ2:} \textit{Does a spherical projection have advantages in supporting network understanding tasks compared to conventional 2D layouts?}

\noindent \textbf{RQ3:} \textit{Does a spherical projection provide perceptual benefits \rev{compared
with arrangements on other 3D topologies, such as a torus?}
}

Before we can answer these questions, we need techniques to create effective layouts of complex network data on a spherical surface and to orient the projections optimally in 2D.

\subsection{Plane, Spherical and Toroidal Stress Minisation}
\label{sec:layoutalgorithms}
The tasks we investigate are cluster understanding tasks and path following.
Network clusters are loosely defined as subsets of nodes within the graph that are more highly connected to each other than would be expected from chance connection of edges within a graph of that density.  More formally, a clustered graph has disjoint sets of nodes with positive \textit{modularity}, a metric due to Newman which directly measures the connectivity of given clusters compared to overall connectivity \cite{newman2006modularity}.  To support cluster understanding tasks we need a layout method which provides good separation between these clusters.

To support path following tasks, we need a layout method which spreads the network out relatively uniformly according to connectivity. This will help minimise crossings between edges.

We follow recent work~\cite{chen2020doughnets,chen2021sa,zheng2018graph} which \rev{adopted a stress minimising approach.  Stress-minimisation is a commonly used variant of a general-purpose force-directed layout} and does a reasonable job of satisfying both of these readability criteria~\cite{huang2009measuring,purchase2002metrics}. The \textit{stress} metric ($\sigma$) for a given layout of a graph with $n$ nodes in a 2D plane is defined (following Gansner et al.\ \cite{gansner2004graph}) as:

$$
    \mathit{\sigma}_\mathrm{plane} = \sum_{i=1}^{n-1} \sum_{j=i+1}^n w_{ij} (\delta_{ij} - d_{ij})^2~~~~,~~~~~~~~~d_{ij} = |x_i - x_j|
$$

\noindent where: $\delta_{ij}$ is the ideal separation between the 2D positions ($x_i$ and $x_j$) of a pair of nodes $(i,j)$ taken as the all-pairs shortest path length between them; $d_{ij}$ is the actual distance between nodes $i$ and $j$ (in the plane this is Euclidean distance); and $w_{ij}$ is a weighting which is applied to trade-off between the importance of short and long ideal distances, we follow the standard choice of $w_{ij} = 1/d_{ij}^2$.

We follow previous recent work by Perry et al.~\cite{perry2020drawing}, in adapting stress-based graph layout to a spherical surface by redefining $d_{ij}$ to arc-length on the sphere surface, or (assuming a unit sphere):

$$
 \mathit{\sigma}_\mathrm{sphere} = \sum_{i=1}^{n-1} \sum_{j=i+1}^n w_{ij} (\delta_{ij} - d_{ij})^2~~~~,~~~~~~~~~d_{ij} = \mathrm{arccos} ( x_i \cdot x_j )
$$

\noindent where $x_i$ and $x_j$ are the 3D vector offsets of nodes $i$ and $j$ respectively from the sphere centroid and $(\cdot)$ is the inner product.  For the layout to be reasonable, the ideal lengths $\delta$ must be chosen such that the largest corresponds to the largest separation possible on the unit sphere surface, which is $\pi$.  Thus, we set the ideal length of all edges to $\pi/\mathit{graph diameter}$.

The other layout against which we compare is a projection of a 3D torus, which, as discussed in \autoref{sec:background} has recently been shown to provide better separation between clusters than a flat (conventional) 2D layout by Chen et al.~\cite{chen2021sa}.  We use the same layout method which is also based on stress in the 2D plane but which, for each node pair, requires selecting the stress term from the set $A$ of 9 possible torus adjacencies for that pair which contributes the least to the overall stress:

$$
 \mathit{\sigma}_\mathrm{torus} = \sum_{i=1}^{n-1} \sum_{j=i+1}^n w_{ij} \mathrm{arg~min}_{\alpha\in A} (\delta_{ij} - d_{ij\alpha})^2 , d_{ij\alpha} = |x_i - x_{j\alpha}|
$$

While Perry et al.\ follow the multi-dimensional scaling literature in using a \textit{majorization} method to minimise $\sigma_\mathrm{sphere}$, we follow Chen et al.~\cite{chen2021sa} in using the stochastic pairwise gradient descent approach developed by \cite{zheng2018graph} which can be adapted straightforwardly and effectively to obtain layout for all of $\sigma_\mathrm{plane},\sigma_\mathrm{sphere},\sigma_\mathrm{torus}$.

Note that the layouts which result from minimising $\sigma_\mathrm{plane}$ and $\sigma_\mathrm{torus}$ are already in 2D.  For the spherical layout obtained by minimising $\sigma_\mathrm{sphere}$, we use either \torthographic{} or \tequalearth{} projections to generate the stimuli for our study. \rev{The detailed pseudocode of our algorithms are available in \url{https://github.com/Kun-Ting/gansdawrap}}.

\subsection{Auto-pan Algorithms}
\label{sec:autopan}

\begin{figure*}[t]
    \centering
     \subfigure[\torthographic{} - No pan]{
    \includegraphics[width=0.4\textwidth]{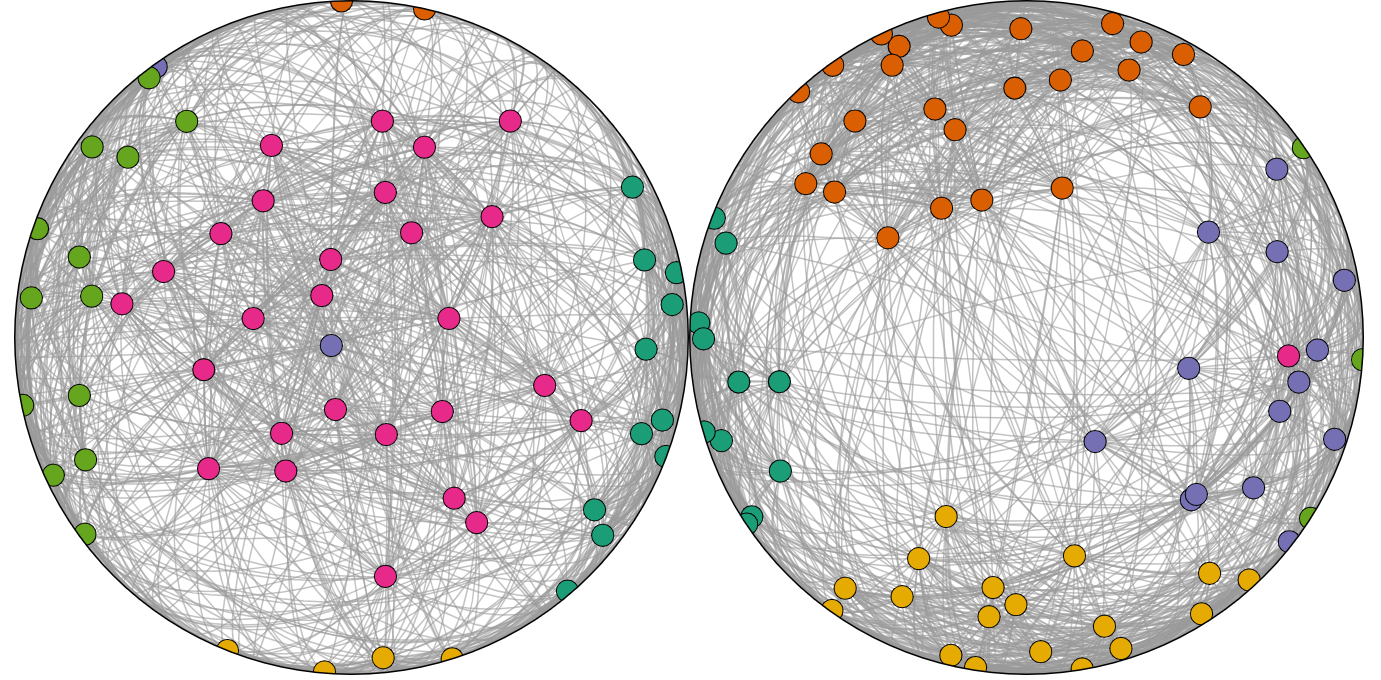}
    }
    \subfigure[\torthographic{} - Best pan]{
    \includegraphics[width=0.4\textwidth]{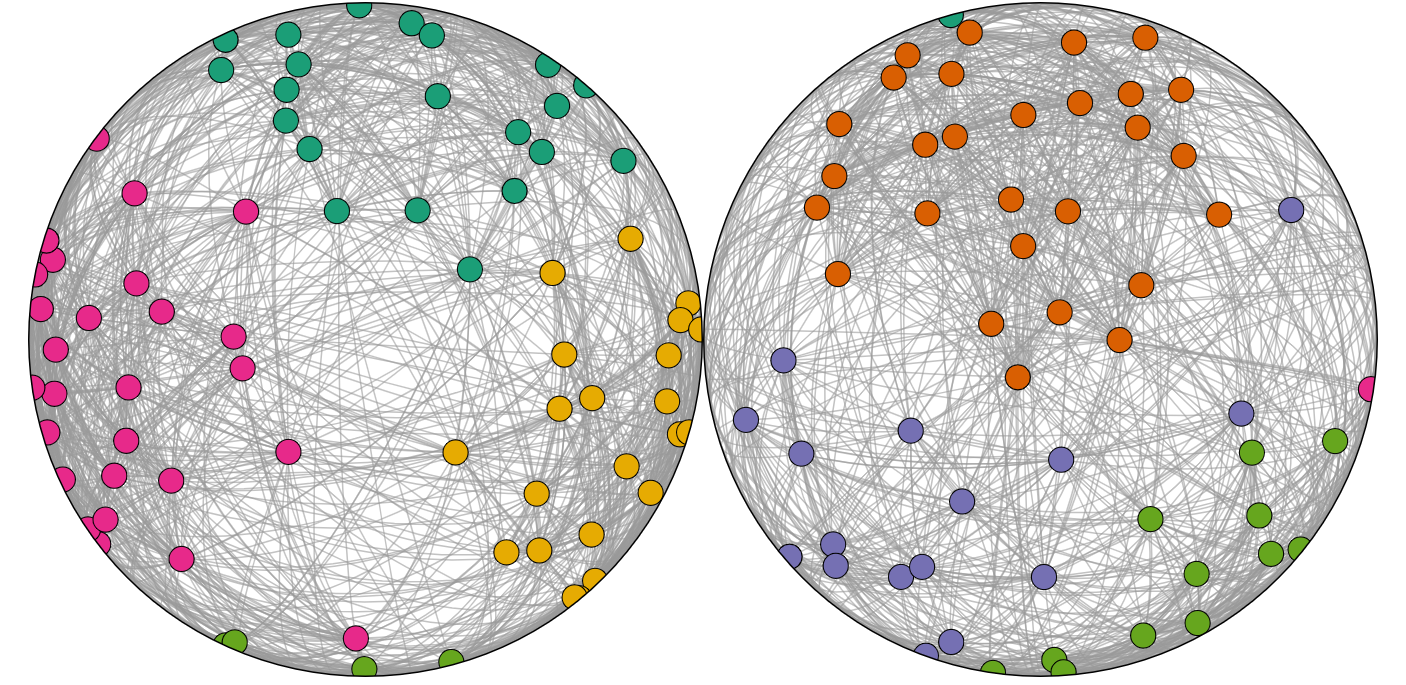}
    }
    \subfigure[\tequalearth{} - No pan (left), edge pixel mask (inset)]{
    \includegraphics[width=0.3\textwidth]{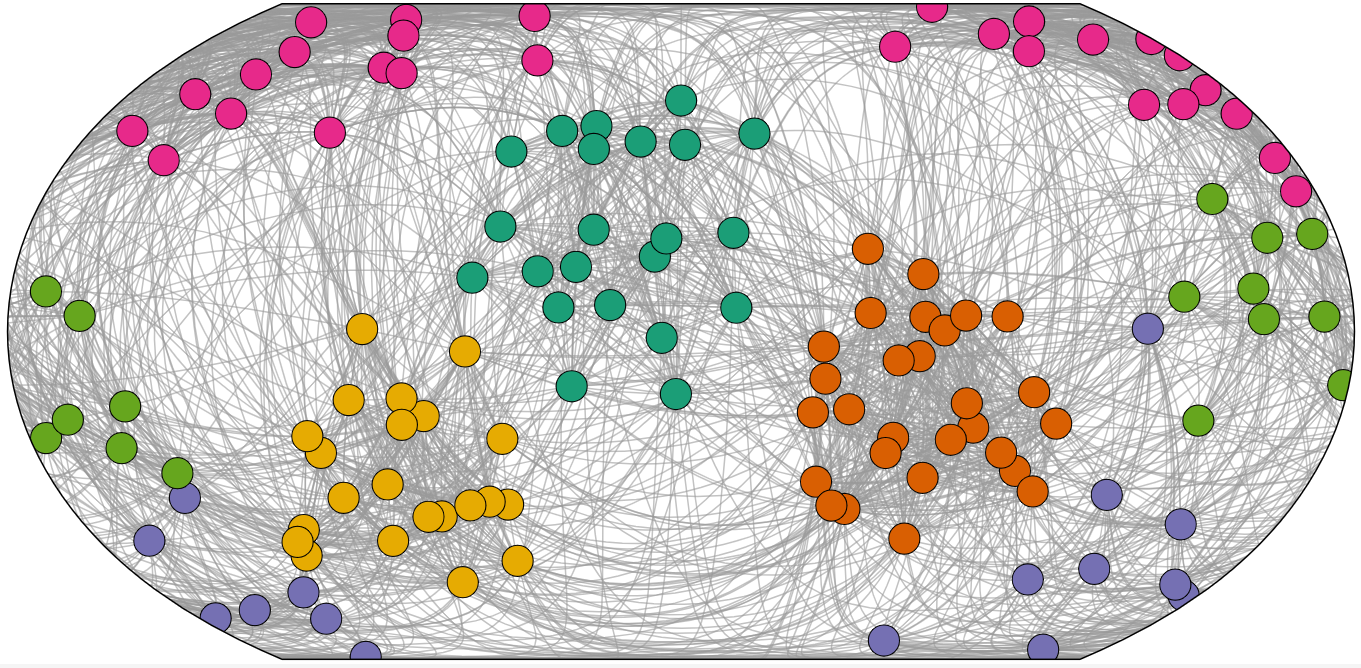}
    \includegraphics[width=0.15\textwidth,trim=1cm 0cm -1cm 0cm ]{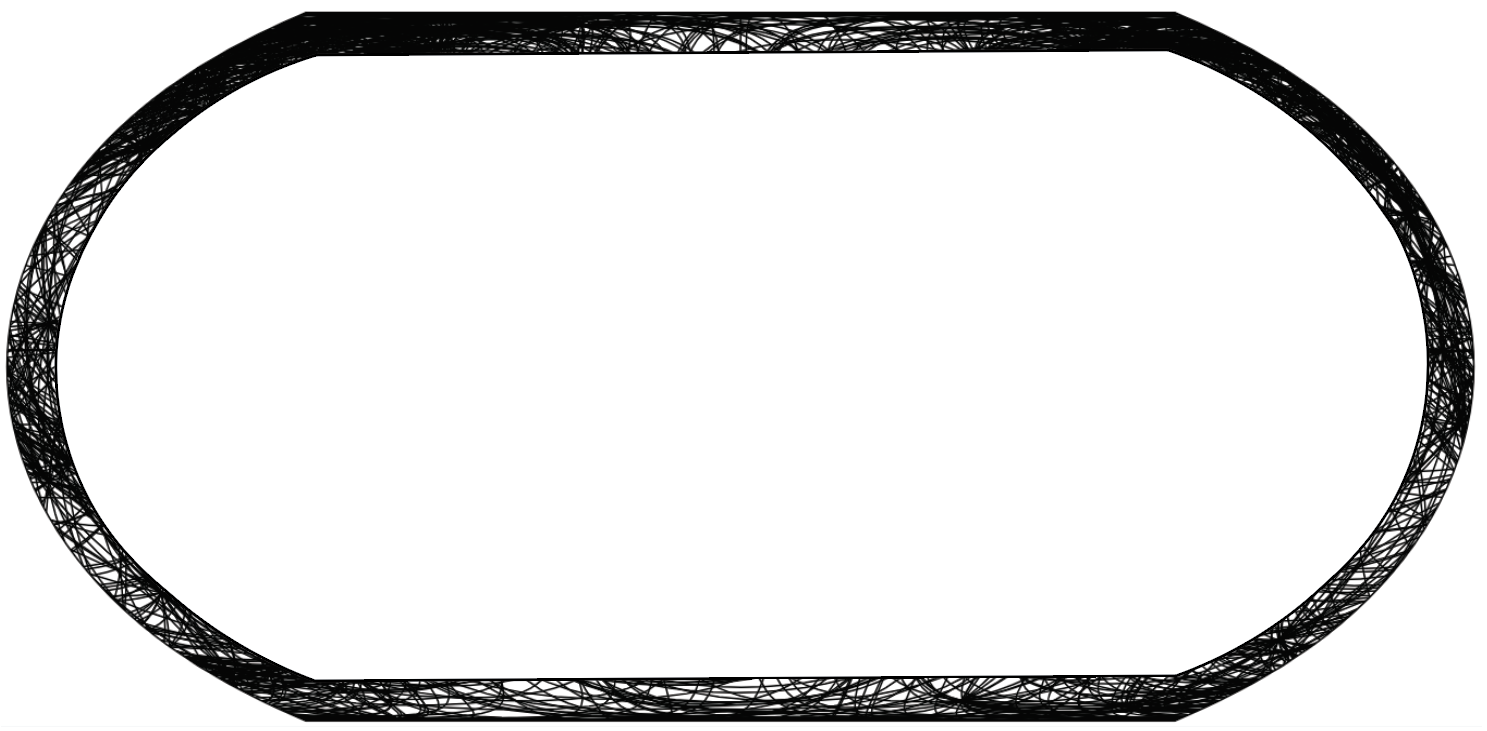}
    }
    \subfigure[\tequalearth{} - Best pan (left), edge pixel mask (inset)]{
    \includegraphics[width=0.3\textwidth]{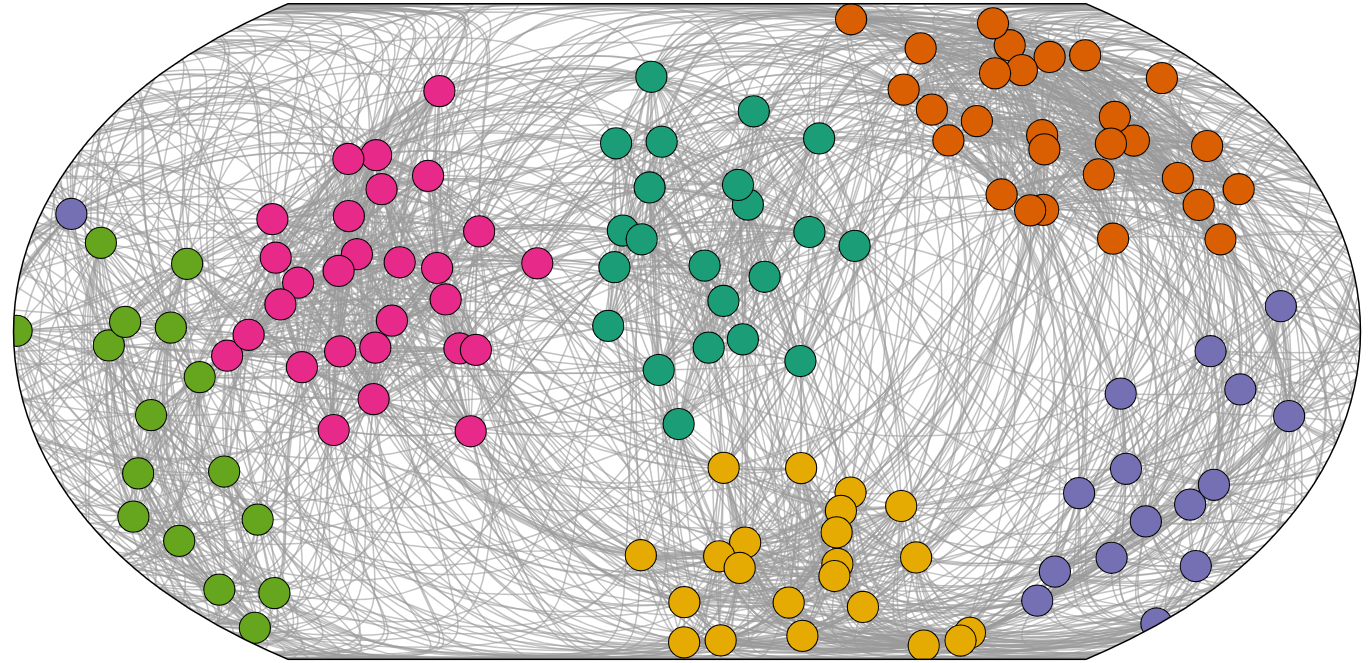}
    \includegraphics[width=0.15\textwidth,trim=1cm 0cm -1cm 0cm]{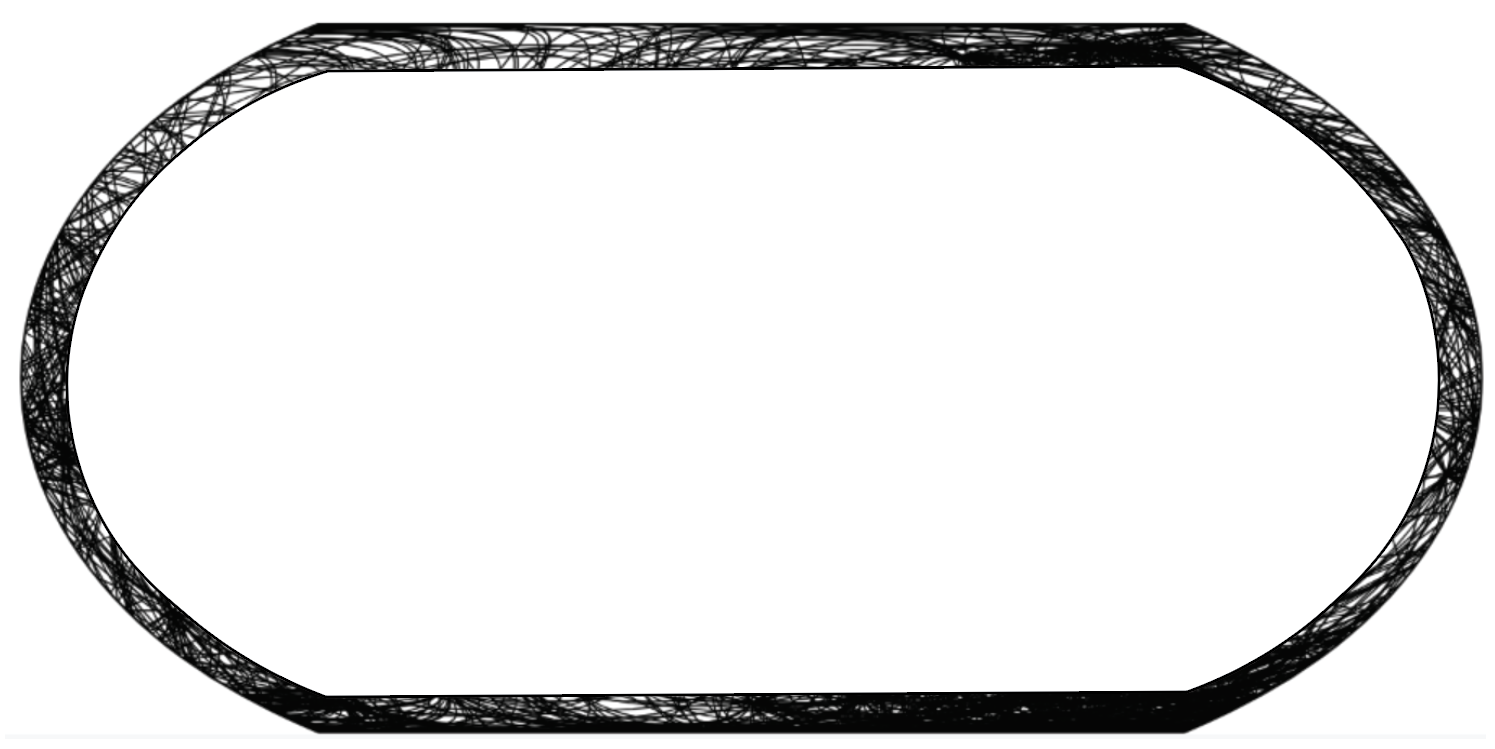}
    }
    \Description[Example of automatic panning]{This diagram shows examples of automatic panning. It shows before and after applying our auto-pan algorithms of a graph with six clusters differentiated by colour. Without auto-pan, clusters can be split across the hemispheres in \torthographic{} (a) or at the boundaries in \tequalearth{} (c).  Auto-pan reduces the number of wrapped edges and thereby brings the clusters together (b) and (d).}
    \caption[]{
    Before and after demonstration of our auto-pan algorithms of a graph with six clusters differentiated by colour (note: study graphs did not have colour).
    Without auto-pan, clusters can be split across the hemispheres in \torthographic{} (a) or at the boundaries in \tequalearth{} (c).  Auto-pan reduces the number of wrapped edges and thereby brings the clusters together (b) and (d).
    }
    \label{fig:autopan}
    \vspace{-1em}
\end{figure*}

For toroidal network layout, Chen et al.\ introduced an algorithm to automatically pan the toroidal layout horizontally and vertically to minimise the number of edges which wrap around at the boundaries \cite{chen2021sa}.  Spherical projections can also suffer when too many edges are split across the boundaries.  Furthermore, edges are more distorted in spherical projections when they are near the edges.  Therefore, for fair comparison with toroidal layouts, it was necessary to find a method to auto-rotate the sphere to reduce the numbers of such edges.  However, while the toroidal auto-panning algorithm can be done with horizontal and vertical scans (linear time in the number of edges), the spherical layout does not permit such a trivial search algorithm.  We therefore develop heuristics to perform auto-rotate for the spherical projections.  For both, we choose a simple stochastic method of randomly selecting a large number \rev{(e.g., 1000 iterations)} of three-axis spherical rotation angle triples $(\lambda,\phi,\gamma)$ and choosing the triple for which edges crossing (or near) boundaries is minimised.

For \torthographic{} projection this crossing number is trivial to count precisely.  Simply, for all pairs of nodes if the nodes are not on the same face, then they must cross a boundary.  A suboptimal \torthographic{} projection rotation, and the result of autorotation to minimise this crossing count is shown in \autoref{fig:autopan}(a) and \autoref{fig:autopan}(b), respectively.

For \tequalearth{} projection the edges are curved and so determining those that cross the boundary for a given geo-rotation is more complex.  Further, in this projection edges near the periphery are significantly more distorted than those near the centre, so even visually determining if an edge that comes close to the boundary continues on the same side of the projection or wraps around to the other side is not easy.  Thus, instead of counting boundary crossings, we penalise all edges which come close to the boundaries.  To compute the penalty we analyse the periphery of a monochrome bitmap of the projected edge paths.  The penalty is then simply the number of black pixels.  Example masked bitmaps are shown for sub-optimal and more-optimal rotations of an \tequalearth{} projection in \autoref{fig:autopan}(c) and \autoref{fig:autopan}(d), respectively.

\subsection{Automatic Panning Results}
\rev{We conducted a small empirical analysis of the Auto-Pan algorithm to assess the numbers of links wrapped for \torthographic{} and the number of pixels (higher values indicate less wrappings) at the boundary of \tequalearth{} projections compared to 10 random rotations. Across 10 study graphs (\dsmalleasy{}, \dsmallhard{}) used in our cluster understanding tasks, we found for \torthographic{}, the mean crossing count for random rotations was 262.16. With automatic panning, this number was improved by $25.6\%$ and was reduced to 208.7.  For \tequalearth{}, our automatic panning increased number of pixels at the boundary region by $12.1\%$ (Section A.2: Table 2).  We found these auto-pan algorithms resulted in a noticeable improvement in keeping clusters from being separated, as evidenced in \autoref{fig:autopan}.
}


\section{Study 2: network projection readability}
\label{sec:networkstudy}

In Study 1, we found \rev{\tequalearth{}, had advantages in terms of error rate, time, and subjective user feedback. \torthographic{}, despite being an interrupted projection, benefited more from interaction than \tmollweide{} and performed well in terms of error rate for distance and direction tasks (\autoref{sec:mapstudydiscussion}).}
Based on these findings, we chose interactive \tequalearth{} and \torthographic{}, to understand their performance in visualising networks. 
We compare these spherical projections to standard flat graph layout (\tnodelink{}) and a projection of a \ttorus{} geometry~\cite{chen2021sa}. 
Our study was run in a similar way to Study 1 and included \rev{a new set of} 96 participants through the Prolific platform.


\subsection{Techniques \& Setup}
The techniques in our study are \tnodelink{}, \ttorus{}, \tequalearth{}, and \torthographic. Layouts are computed as described in \autoref{sec:layout}.
All the techniques support interactive panning except for \tnodelink{} which does not wrap around.
The area of the rectangular bounding box of each technique condition is the same. For \tnodelink{} and \ttorus{}, \rev{the} resolution is fixed at $650 \times 650$ pixels, and for \tequalearth{} and \torthographic{}, \rev{the} resolution is fixed at $900 \times 317$ pixels.
For \tclusteridentification{}, we did not color clusters to not reveal any graph structure.

\subsection{Tasks}
\begin{figure*}
    \centering
    \includegraphics[width=0.9\linewidth]{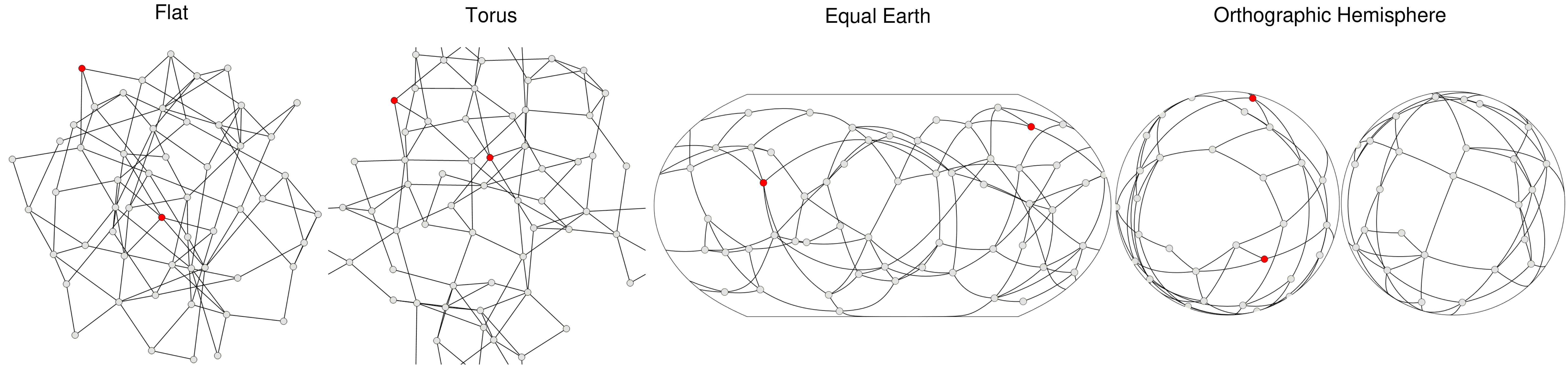}
    \Description[]{This diagram shows an example of one study graph laid out using 4 techniques described in Section 5.1 for shortest path number task. The shortest path length is 3 in this example. Participants were provided with interactive panning for spherical and toroidal layouts to explore the network.}
    \caption[]{\rev{Example of one study graph laid out using 4 techniques described in \autoref{sec:layoutalgorithms} for \tshortestpath{}. The shortest path length is 3 in this example. Participants were provided with interactive panning for spherical and toroidal layouts to explore the network.}}
    \label{fig:graphtasks}
\end{figure*}
We selected two representative network visualisation tasks, inspired by existing work comparing 2D layout of 3D surface topology~\cite{chen2021sa, chen2020doughnets} and task taxonomies~\cite{lee2006tasktaxonomy, saket2014group}.

\noindent \textbf{\tclusteridentification:} \textit{Please count the number of clusters in this graph.} Participants answered through radio buttons: choices ranged from 1 to 10. We created a quality control trial for each condition with two clusters with clearly marked boundaries to assess participants' attention.  Participants who did not answer these trials correctly were excluded. \rev{An example of 5 clusters from \dlargeeasy{} graphs (\autoref{sec:networkdatasets}) using 4 different layouts is shown in \autoref{fig:teaser}-Bottom.} 

\noindent \textbf{\tshortestpath:} \textit{What is the shortest path length between the red nodes?}: Participants were required to count the smallest number of links between two red nodes. Radio buttons allowed them to answer between 1 to 6. Again, we created a quality control trial for each layout condition, with shortest path length two and links on the path highlighted in red. \rev{An example of \deasy{} graphs  (\autoref{sec:networkdatasets}) using 4 different layouts is shown in \autoref{fig:graphtasks}.} 

\subsection{Data Sets}
\label{sec:networkdatasets}
We prepared a separate graph corpus for each task, full details and stimuli are presented in our supplementary material. For \tclusteridentification, we use graphs from Chen et al.~\cite{chen2021sa}, generated using algorithms designed to simulate real-world community structures in graphs~\cite{brandes2003experiments,fortunato2010community}. 
Graphs are grouped by two variables: difficulty in terms of graph modularity~\cite{newman2006modularity} (\deasy: modularity=0.4, \dhard: modularity=0.3) and size (2 levels: \dsmall{}: 68-80 nodes, 710-925 links, and \dlarge{}: 126-134 nodes, 2310-2590 links). The number of clusters is between 4 and 7. 
%
For \tshortestpath{} the clustered graphs were too dense, so we generated sparser graphs using scale-free models~\cite{barabasi1999emergence,watts1998collective}.
We chose graphs with two levels of density (\deasy{}: 0.075, and \dhard{}: 0.11) with 50 to 57 nodes.
The shortest path length varied between 1 and 4.

\subsection{Hypotheses}

Hypotheses were pre-registered with the Open Science Foundation:~\url{https://osf.io/equhp}.

\noindent \textbf{H2-1}: \textit{\tequalearth{} and \ttorus{} have better task effectiveness for \tclusteridentification{} (in terms of time and error) than \torthographic} \rev{(RQ1, RQ3)} \textit{or \tnodelink{}} \rev{(RQ2)}. While \torthographic{} performed well in Study 1, our pilot studies for \tclusteridentification{} revealed that cuts and distortion of clusters at the borders made them hard to count. \rev{For (RQ2), the inspiration was based on prior cluster readability studies~\cite{chen2021sa}.} 

\noindent\textbf{H2-2}: \textit{Participants will prefer \tequalearth{} and \ttorus{} to \torthographic{}} \rev{(RQ1, RQ3)} \textit{or \tnodelink{}} \rev{(RQ2)} \textit{for \tclusteridentification.} Our early pilots indicating this preference --- perhaps for the same reasons as above \rev{and inspiration from existing studies~\cite{chen2021sa}}.
    
\noindent\textbf{H2-3}: \textit{\ttorus{} has better task effectiveness \rev{(in terms of time and error)} for \tshortestpath{} than \tnodelink{} \tequalearth{} or \torthographic{}} \rev{(RQ3)}. This assumption is based on pilot studies and prior studies indicating curved links might hamper path tracing tasks~\cite{du2017isphere, xu2012user}.
    
\noindent\textbf{H2-4}: \textit{Participants will prefer \ttorus{} to \tnodelink{}, \tequalearth{} or \torthographic{}} \rev{(RQ3)} \textit{for \tshortestpath{}.} Again, it was assumed due to distortion of links.

\subsection{Experimental Design}

We use a within-subjects design for each task with 4 techniques. Each participant was randomly assigned one of the tasks by the experimental software. For \tclusteridentification, we used 2 levels of difficulty (Easy, Hard) $\times$ 2 sizes (Small, Large) $\times$ 5 repetitions. We randomly inserted one additional quality control trial to each layout condition. Each recorded trial had a timeout of 20 seconds to prevent participants from trying to perform precise link counting. This leaves us with a total of 80 recorded trials per participant. We counterbalanced the order of the techniques using a full-factorial design. The order of each level of difficulty and size in each technique was the same: \dsmalleasy{}$\rightarrow$\dlargeeasy{}$\rightarrow$ \dsmallhard{}$\rightarrow$\dlargehard{}. The order of trials for each technique within each level was randomised. 

For \tshortestpath, we used 2 levels of difficulty (Easy, Hard) $\times$ 8 repetitions. There were 2 repetitions of each shortest path length per level. \rev{One additional quality control trial was added for each layout condition.} This leaves us with a total of 64 trials per participant. Each recorded trial had a timeout of 30 seconds, informed by pilot studies. The order of each level of difficulty in each technique was the same: \deasy{}$\rightarrow$ \dhard{}. The order of trials for each technique within each level was randomised. 

\subsection{Participants and Procedures}

Setup and inclusion criteria for participants were the same as for Study 1. 
We recorded 96 participants (37 female, 59 male, age range [18,50]) who passed the attention check trials, completed the training and recorded trials. This comprised 2 fully counterbalanced blocks of participants (24 $\times$ 2 $\times$ 2 tasks). They ranked their familiarity with network diagrams as: 9 often seeing network diagrams; 62 occasionally; and 25 never. 

For \tclusteridentification{}, each participant was presented with a tutorial explaining the concept of network clusters. For \ttorus \tequalearth, and \torthographic, animated videos were given to demonstrate the interactive panning or rotation. Each participant then completed training trials,  similar to the recorded trials. 
For \tshortestpath, each participant was presented with a tutorial explaining the concept. For \ttorus, \tequalearth, \torthographic, animated videos were given to demonstrate the interactive panning or rotation. 
Training trials were similar to the recorded trials. 

Specific instructions given for each task is available in the supplementary material\footnote{An online demonstration of the study is available: \url{https://observablehq.com/@kun-ting/gansdawrap}}.

\subsection{Dependent Variable and Statistical Analysis Methods}

We recorded task-completion time (\mtime), task-error (\merror), and subjective \textit{confidence} as \mpref{} (the smaller the more confident). We calculated \merror{} as the normalised absolute difference between the correct answer and response. We used the same statistical analysis methods \rev{and standardised effect sizes} from the first study (\autoref{sec:study-1-stats}).

\begin{figure*}
    \centering
    \includegraphics[width=\linewidth]{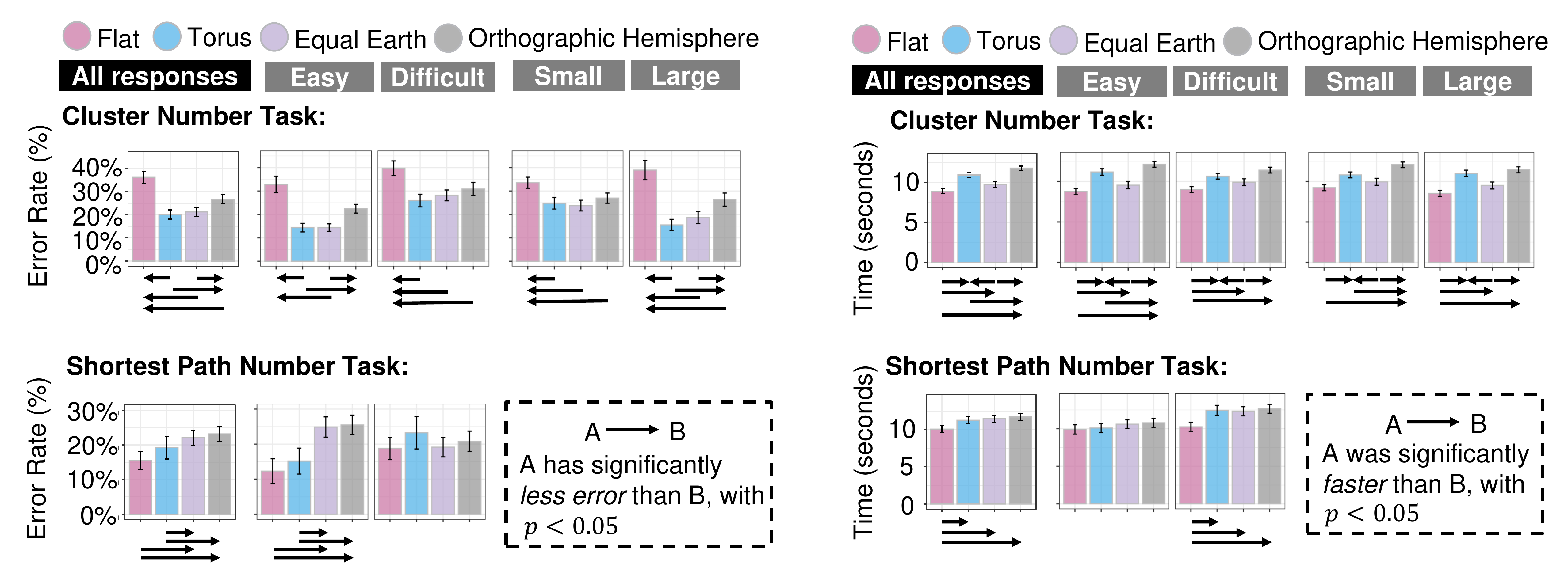}
    \Description[]{This diagram shows an the results of Study 2 in terms of error rate (left) and Time (right). Error bars are 95\% confidence intervals. Significant differences between projections are shown as arrows.}
    \caption{Error rate (left) and Time (right) results of Study 2. \rev{Error bars are 95\% confidence intervals.} Significant differences between projections are shown as arrows. \rev{Detailed statistical results and effect sizes are available in Section A.2.}}
    \label{fig:study-2-results}
\end{figure*}


\begin{figure}
    \centering
    \includegraphics[width=\linewidth]{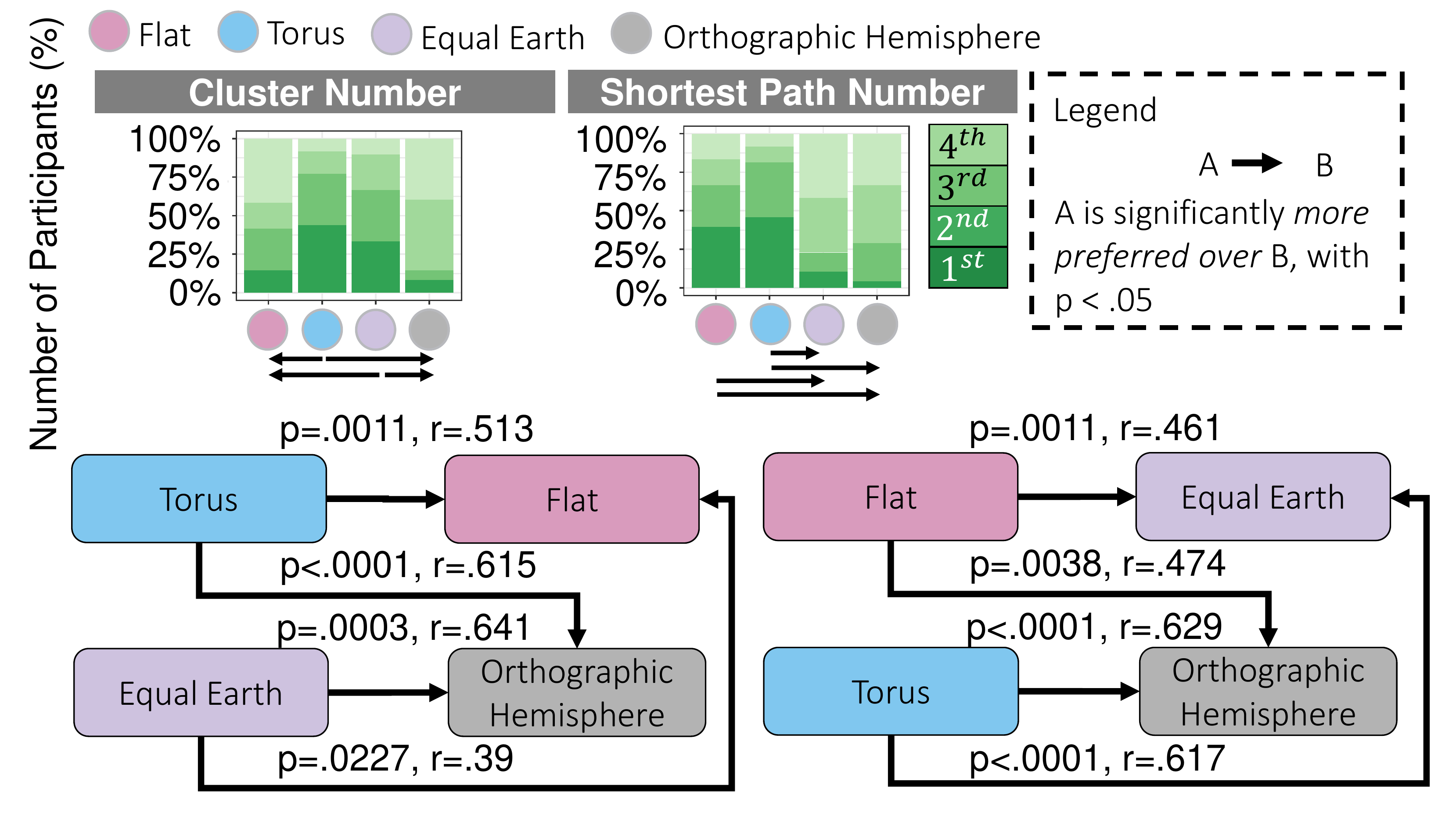}
    \Description[Legend]{This diagram shows user confidence ranking of each condition for two tested tasks. Arrows indicate statistical significance with $p<0.05$. Effect size results of Cohen's r~\cite{cohen2013statistical} are presented along with the arrows.}
    \caption{User confidence ranking of each condition for two tested tasks. Arrows indicate statistical significance with $p<0.05$. \rev{Effect size results of Cohen's r~\cite{cohen2013statistical} are presented along with the arrows.}}
    \label{fig:study-2-ranking-results}
     \vspace*{-6pt}
\end{figure}

\subsection{Key Findings and Discussion}
\label{sec:networkstudykeyfinding}
\rev{We report on the most significant findings for \tclusteridentification{}, \tshortestpath{} visually in \autoref{fig:study-2-results} and \autoref{fig:study-2-ranking-results}. Detailed statistical results and pairwise effect sizes are provided in Section A.2}.


\textbf{\rev{For \tclusteridentification{}, \tequalearth{} and \ttorus{} performed equally well and \tequalearth{} was slightly faster than \ttorus{}, while they both significantly outperformed \torthographic{} in terms of \merror{}, \mtime{}, and \mpref{}}.}
\rev{The poor performance of \torthographic{} compared with \tequalearth{} and \ttorus{} was found statistically significant with moderate and large effect sizes for \merror{} (All, Easy, Large), \mtime{} (Easy, Large), and large effects for \mpref{} (\autoref{fig:study-2-results}-top, \autoref{fig:study-2-ranking-results}, Section A.2: Figure 15, Figure 16, and Figure 17)}, \rev{confirming \textbf{H2-1, H2-2} for (RQ1, RQ3)}.  
Surprisingly, while results from Study 1 (\autoref{sec:mapstudydiscussion}) showed that \rev{interactive \torthographic{} tended to be more accurate than \tequalearth{} for distance comparisons and not worse than any other projections for area comparisons}, it turned out to be significantly worse than \tequalearth{} and \ttorus{} for reading \tclusteridentification. 

\rev{While it has the advantage of being a straightforward mapping from the sphere,} \torthographic{} is an interrupted and non-area-preserving projection.  Therefore, we conjecture that this discontinuity caused participants to struggle to make out cluster boundaries \rev{as compared with continuous representations in \ttorus{} and \tequalearth{}. Furthermore, there is no link distortion in \ttorus{}. This was supported by participants feedback, e.g., \emph{``The `Equal Earth' method felt much easier to distinguish every [individual] cluster.''} (P20), and \emph{``Orthographic Hemisphere uses 2 maps so it is much more difficult to interpret than the others [...] making it harder to isolate clusters.''} \rev{(P12)}.} More quotes can be found in  Section A.2.1.

\rev{We also note that although being statistically significant, the small effect sizes indicate that \tequalearth{} is slightly faster than \ttorus{}~\cite{helske2021can,cockburn2020threats}}. 

\textbf{\tequalearth{}, \torthographic{} and \ttorus{} significantly resulted in less error than \tnodelink{}. \rev{\tequalearth{} and \ttorus{} were significantly preferred over \tnodelink{} but \ttorus{} and \torthographic{} took longer time than \tnodelink{}.}}
\rev{This was found statistically significant with moderate and large effect sizes (\autoref{fig:study-2-results}-top, \autoref{fig:study-2-ranking-results}, Section A.2: Figure 15, Figure 16, and Figure 17), leading us to reject \textbf{H2-1} for (RQ2) and confirm \textbf{H2-2} for (RQ2).} These results provide strong evidence that with automatic panning and interaction, participants were able to better identify the high-level network structures using spherical projections. 
They also confirm Chen et al.'s results where toroidal layouts with automatic panning significantly outperformed \tnodelink{} in terms of error for cluster understanding tasks~\cite{chen2021sa}. Some participants explicitly mentioned that good separation of clusters \rev{in continuous surfaces such as \tequalearth{} and \ttorus{}} helped understanding \rev{(Section A.2.1)}. 

Participants also mentioned the automatic panning helped them ``see'' without the need to interact e.g., \emph{``[with \tequalearth,] even though if we move the picture [it] is very hard to understand the clusters. [It] turned out to be easiest to find the clusters (by not moving the picture).''}  \rev{(P37)}. 

\textbf{For \tshortestpath{}, \tnodelink{} and \ttorus{} \rev{performed equally well}, both significantly outperforming \tequalearth{} and \torthographic{} in terms of \merror{} \rev{and \mpref{}, while \tnodelink{} is faster than all representations.}} \rev{We found \tnodelink{} and \ttorus{} have a lower error rate than \tequalearth{} and \torthographic{} for all responses (moderate effects) and Easy (large effects) (\autoref{fig:study-2-results}-bottom, Section A.2:  Figure 15 and Figure 16). There is a strong evidence that \ttorus{} is more preferred (large effects) over \tequalearth{} and \torthographic{}, while \tnodelink{} is more preferred (moderate effects) over \tequalearth{} and \torthographic{} (\autoref{fig:study-2-ranking-results} and Section A.2: Figure 17). \tnodelink{} was the fastest for All Responses (small and moderate effects) and Hard (moderate and large effects). We therefore accept \textbf{H2-3, H2-4} for (RQ3) but reject \textbf{H2-3, H2-4} for \tnodelink{}.}

\rev{We conjecture that} although \ttorus{} involves broken links across the boundaries, it appears similar to \tnodelink{} using straight links
while the distortion of paths in \tequalearth{} might hamper path tracing tasks, \rev{as participants mentioned }\emph{``Flat surface is easier to read, it helps sometimes when you can also move it. Earth-like is just hard to use, especially when it's an equal globe.''} (P9), and \emph{``the warping in [\tequalearth{}] made my eyes hurt a little therefore its in 4th place, and the torus being the most straight forward gets 1st with [\tnodelink{}] in second as they're very similar.''} (P13). On the other hand, \torthographic{} has less link distortion but the interruption between two hemispheres and the strong perspective distortion near the boundary of the maps may make participants confused, \rev{supported by participants' feedback (Section A.2.1)}. 

Although automatic panning (\autoref{sec:autopan}) provides some benefits for reducing split of clusters across the boundaries for \tclusteridentification{}, it seems it has less benefits for \tshortestpath{}, as participants mentioned they still need to position the image to see the full path to identify the shortest path connecting the nodes within the time limit (Section A.2.1).


Overall, these findings suggest that \ttorus{} presents a general solution being not only less error prone than \tnodelink{} or \torthographic{} for \tclusteridentification{}, but also comparable to \tnodelink{} for \tshortestpath{}.

\subsection{Limitations}
\label{sec:study-2-threats}


\rev{A limitation of our study is that we only tested with one layout algorithm. While many different algorithms exist for \tnodelink{} and it is possible that other layout algorithms could be adapted to spherical and torus embeddings, doing so is not necessarily trivial. 
Also, it should be noted that there are algorithms which can optimise layout for a known set of clusters (i.e.\ where the cluster labelling is known in advance) ~\cite{meulemans2013kelpfusion, collins2009bubble,gansner2009gmap,hu2010visualizing}. However, for the tasks tested in this paper we do not preidentify the clusters but rather leave cluster identification as the user task. 
Similar layouts~\cite{dwyer2008topology} have been used before for cluster understanding tasks on torus layouts~\cite{chen2020doughnets,chen2021sa}.}  

\section{Conclusion and Future Work}
    
The overwhelming finding in Study 1 that interactive panning led to significantly lower error rate \rev{and more preferred over standard map projections} across all tasks tested suggests that such panning should be routinely provided in online world maps, for example in education or in reporting of world events, climate patterns, and so on.  While less clear, our results indicated that \tequalearth{} was the best performing continuous projection, while the straight-forward \torthographic{} was an effective hemispheric projection. \rev{Our results also indicate that even with interaction, \tequirectangular{} may be a poor choice for comparing areas and estimating directions.}

Our Study 2 results also confirm the benefits of topologically closed surfaces, such as the surface of a torus or sphere, when using node-link diagrams to investigate network structure.  \rev{All of \tequalearth{}, \torthographic{}, and \ttorus{}} outperformed \tnodelink{} for cluster understanding tasks.  While the spherical projections impeded path following tasks, it seems \ttorus{} may be a good general solution, being as accurate as \tnodelink{} for path following.
Although they have been explored in research, 2D projections of toroidal and spherical network layouts are rarely seen in practice.  This may be because until recently effective layout and projection methods for such geometries were not easily available.  We intend to make all our algorithms and extensions of existing open-source tools available for easy consumption in web applications via GitHub and {\tt npm} packages.

A limitation of our work is that there are many more spherical map projections than those evaluated here, although we tried to select the most representative techniques.  Further work may extend such evaluation to other projection types.  Another interesting topic for future investigation would be investigating spherical and torus projections of other types of abstract data representation, such as multi-dimensional scaling techniques of high-dimensional data.  \rev{Another family of interactive techniques for exploring graphs involve applying spatial distortion around regions of interest to achieve a kind of ``structure aware zooming'', e.g. \cite{wang2018structure}.  It would be interesting to compare the efficacy of these techniques against interactive sphere and torus projections.}

\begin{acks}
\textit{\rev{We thank Bernhard Jenny, Lonni Besançon, Sarah Schöttler, Hong Gui, Mengxing Li, Ishwari Bhade, Umair Afzal, Sunny Singh for helpful discussion about the study design, anonymous participants for user study feedback, and our reviewers for helpful suggestions to improve this manuscript. We thank Stay Healthy and Monash eSolution crew for their technical support. This research is supported in part by Monash FIT Graduate Research Candidature Funding Scheme.}}
\end{acks}

\bibliographystyle{ACM-Reference-Format}
\bibliography{main}

\appendix
\section{Study Results}
\label{sec:supplementary}
In the following, significance values are reported for $p < .05 (*)$, $p < .01 (**)$, and $p < .001 (***)$, respectively, abbreviated by the number of stars in parenthesis. 
\subsection{Study 1: Map Projection Readability}
\label{sec:supplestudy1}
\textbf{For \tdistancecomparison{}}, we found \finteraction{} ($***$) and \fmap{} $\times$ \fdifficulty{} ($***$) both had a significant effect on \mtime{}. For \finteraction{}, posthoc analysis shows that
\dstatic{} was faster than \dinteractive{} ($***$).
For \fmap{} $\times$ \fdifficulty{}, posthoc analysis shows that in \deasy{}, \tequirectangular{} and \tequalearth{} were faster than \torthographic{} ($**$) and \tmollweide{} ($***$). \thammer{} was also faster than \tmollweide{} ($***$). In \dhard{}, \tequalearth{} was faster than \torthographic{} ($*$) and \tmollweide{} ($***$).

We found \fmap{} had a significant effect on \emph{error rate} in \dhard{} ($*$). \torthographic{} was more accurate than \tequalearth{} ($*$) and \thammer{} ($*$). We also found \dinteractive{} was more accurate than \dstatic{} in all \fmap{} ($***$). 

In \dstatic{} \fmap{}, participants preferred \tequalearth{} over \thammer{} ($*$), \tmollweide{} ($*$), and \torthographic{} ($**$). 

\textbf{For \tareacomparison{}},  we found \finteraction{} ($***$), \fmap{} $\times$ \fdifficulty{} ($***$) and \fmap{} $\times$ \finteraction{} ($**$)  had a significant effect on \emph{time}. For \tareacomparison{}, \dstatic{} was faster than \dinteractive{} ($***$). For \fmap{} $\times$ \finteraction{} ($**$), \dinteractive{} \tequalearth{} was faster than \dinteractive{} \thammer{} ($**$).

We found \fmap{} had a significant effect on \merror{} in \dinteractive{} ($*$), \deasy{}  \dstatic{} ($*$) and \dhard{} \dinteractive{} ($*$). \tmollweide{} was more accurate than \tequirectangular{} ($*$). 
\dstatic{} \tequalearth{} ($*$) and \dstatic{}\tmollweide{} ($*$) were more accurate than \dstatic{} \tequirectangular{} in \deasy{}.
\dinteractive{} \tmollweide{} and \dinteractive{} \dinteractive{} tended to be more accurate than \dinteractive{} \tequirectangular{}, but not statistically significant (with $p=0.09, 0.07$ respectively).
We also found \dinteractive{} was more accurate than \dstatic{} in all \fmap{} (all $***$). 

In \dstatic{} \fmap{}, participants preferred \tequirectangular{} ($***$), \tequalearth{} ($**$) and \torthographic{} ($**$) than \thammer{}.

\textbf{For \tdirectionestimation{}},  we found \fmap{} ($***$), \finteraction{} ($***$) and \fmap{} $\times$ \finteraction{} ($**$) had a significant effect on \emph{time}. For \tdirectionestimation{}, \dstatic{} was faster than \dinteractive{} ($***$).
In \dstatic{} \fmap{}, \tequirectangular{} and \tequalearth{} were faster than \tmollweide{} and \torthographic{} (all $**$).
In \dinteractive{} \fmap{}, \tequalearth{} was faster than \tmollweide{} ($**$).

We found \fmap{} had a significant effect on \emph{error rate} in \dinteractive{}. \torthographic{} was more accurate than \tequirectangular{} ($*$). We also found \dinteractive{} was more accurate than \dstatic{} in all \fmap{} ($***$). 

We did not find significant difference in preference for \dinteractive{} groups.

\subsubsection{Qualitative Feedback} 
\label{sec:supplequalitativefeedback}

\textbf{In continuous projections, \tequalearth{} \rev{performed well in terms of \merror{} for \tareacomparison{}, \mtime{} for \tareacomparison{} and \tdirectionestimation{}, and \mpref{}-\dstatic{} for both \tdistancecomparison{} and \tareacomparison{}. \tequalearth{} was not significantly worse than any other continuous projections.}}
The preference of \dstatic{} \tequalearth{} was supported by our participants' feedback, as they mentioned \emph{``I didn't feel comfortable when the hemispheres were split into two separate circle portions. I preferred [equal earth] when they were a continuous map surface.''} (P18, Distance-\dstatic{}), and \emph{``the map's edges [in equal earth projection] didn't feel as distorted as other maps such as equirectangular.''} (P23, Area-\dstatic{}.) We also identified many positive comments for \tequalearth{}, e.g., \emph{``[equal earth projection] gave me a better sense of global spatial reasoning''} (P34, Distance-\dstatic{}), and \emph{``the equal earth projection was the easiest to visualise as it seemed to have a good balance of not being too distorted without having the difficulty of wrap around visualisations that the hemisphere projections had.''} (P37, Distance-\dstatic{}).

Participants disliked \thammer{} for \dstatic{}-Area. For example, participants mentioned \emph{``I think [equirectangular] felt easier to read my first choice vs what I rate number 5 [for hammer].''} (P17, Area-\dstatic{}) and \emph{``the hammer too oval to make sense of.''} (P19, Direction-\dstatic{}).

\textbf{In hemispheric projections, interaction reduced \merror{} of \torthographic{} to a point that it tended to have a lower error rate than some interactive continuous projections for Distance-Hard and Direction, and not significantly slower than any interactive projections across all tasks.}
\torthographic{} was perceived as less distorted than the other projections due to the ``natural'' orthographic distortion, which is similar to viewing the sphere at infinite distance (e.g.\ as if through a telescope). Furthermore, despite being hemispheric, \tmollweide{}, was found to be less intuitive for some tasks by participants. We conjecture that there is a slight distortion near the edge of two circles which may make it confusing when centring the region of interest, as participants mentioned \emph{``Orthographic hemisphere felt the easiest to compare because it was like looking at a globe, but the strange way of moving the mollewide hemisphere made it confusing, that is why that is my last option.''} (P23, Area-\dinteractive{}), \emph{``The mollweide was a bit difficult to drag and place arrow.''} (P23, Direction-\dinteractive{}.), and \emph{``Orthographic Hemisphere seemed the most accurate to me. The round shape was really useful. The Mollweide Hemisphere was tricky, I didn't like the way the two points had to be on the separate circles for it to be accurate sometimes.''} (P4, Direction-\dinteractive{}). 

However, there were also participants who did not like the hemisphere projections due to the need to inspect two separated spheres and instead they preferred non-hemisphere, e.g., \emph{``I found it a lot easier to identify on the single maps rather than the double because sometimes they were spaced too far away''} (P11, Area-\dinteractive{}), \emph{``it was hard to picture how Mollweide and Orthographic Hemispheres connected.''} (P7, Distance-\dstatic{}), and \emph{``the map layout was easier to see because the targets were all on one map rather than two spheres.''} (P36, Distance-\dinteractive{}).

Meanwhile, it seems many people were good at mentally connecting both hemispheres: \emph{``I found it easier to create comparisons that didn't have a lot of distortion with the option of having the two side-by-side globes.''} (P18, Direction-\dinteractive{}). 

\textbf{Even with interaction, \tequirectangular{} still tended to perform poorly in terms of \merror{} for \tareacomparison{} and \tdirectionestimation{}.}
\dinteractive{} \tequirectangular{} was still found to have a greater distortion near the poles than other projections, as participants mentioned \emph{``The sphere is easiest to visualize in my head, I was thinking about just drawing a line on a ball or other spherical shape. The ones with the biggest distortment on the edges confused me most.''} (P13, Direction-\dinteractive{}) and \emph{``The Mollweide Hemisphere and Orthographic Hemisphere- easier to see a side by side comparison of the areas. The Hammer- less areas of distortion compared to the Equal Earth and Equirectangular. Equal Earth had too much distortion as did the Equirectangular which in my opinion distorted the most.''} (P21, Area-\dinteractive{}).

\begin{figure*}
    \centering
    \includegraphics[width=.68\textwidth]{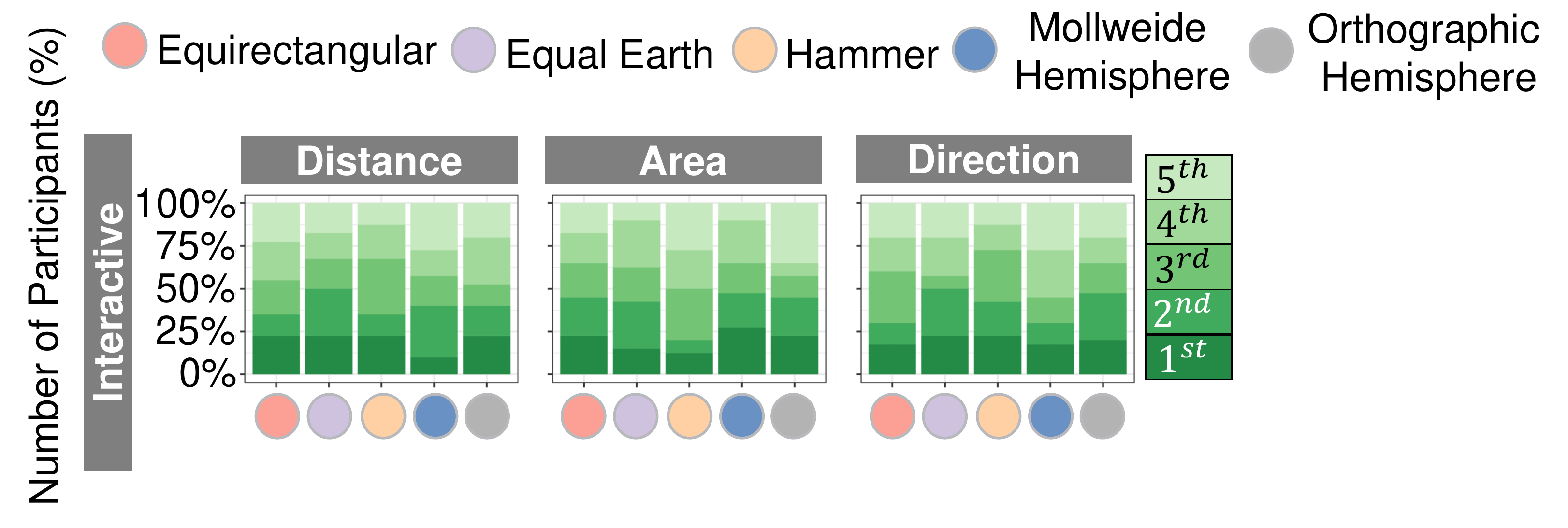}
    \Description[Legend]{}
    \caption{Study 1: Subjective user rank of map projections within the \emph{interactive} group for three tested tasks. Higher rank indicates stronger preference. Interaction makes the preference between \emph{interactive} projections similar (not statistically significant). Effect sizes can be found in \autoref{fig:study-1-ranking-results-effect-size}.}
    \label{fig:study-1-ranking-results-interactive}
\end{figure*}

\begin{figure*}
    \centering
    \includegraphics[width=.68\textwidth]{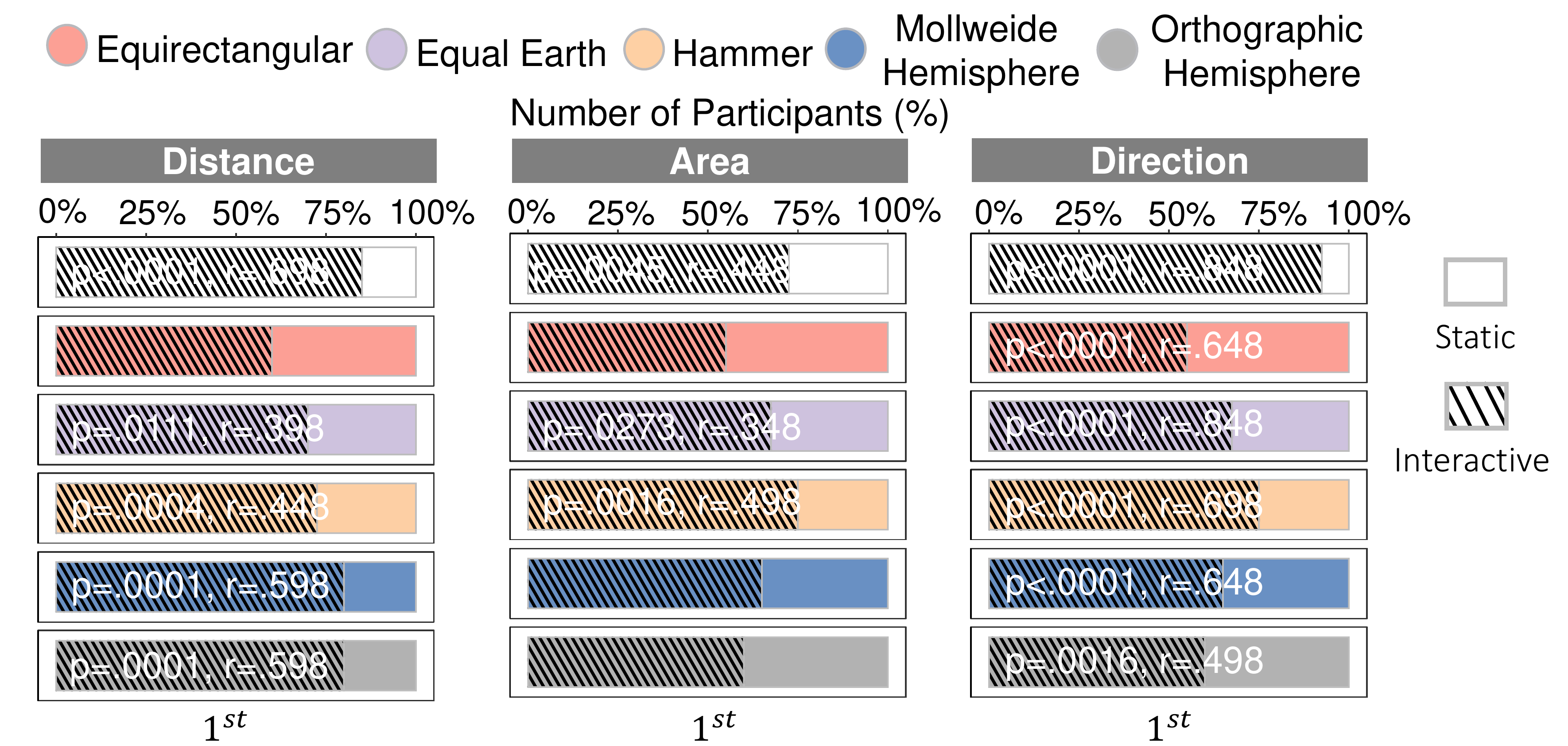}
    \Description[Legend]{}
    \caption{Study 1: User preference rank of interactivity for overall (top) or within each map projection. Statistical significance results with $p<0.05$ and effect sizes of Cohen's r are shown. x-axis shows the number of participants in percentage.}
    \label{fig:study-1-interactivity-ranking-results}
\end{figure*}

\begin{figure*}
   \centering
    \includegraphics[width=1\textwidth]{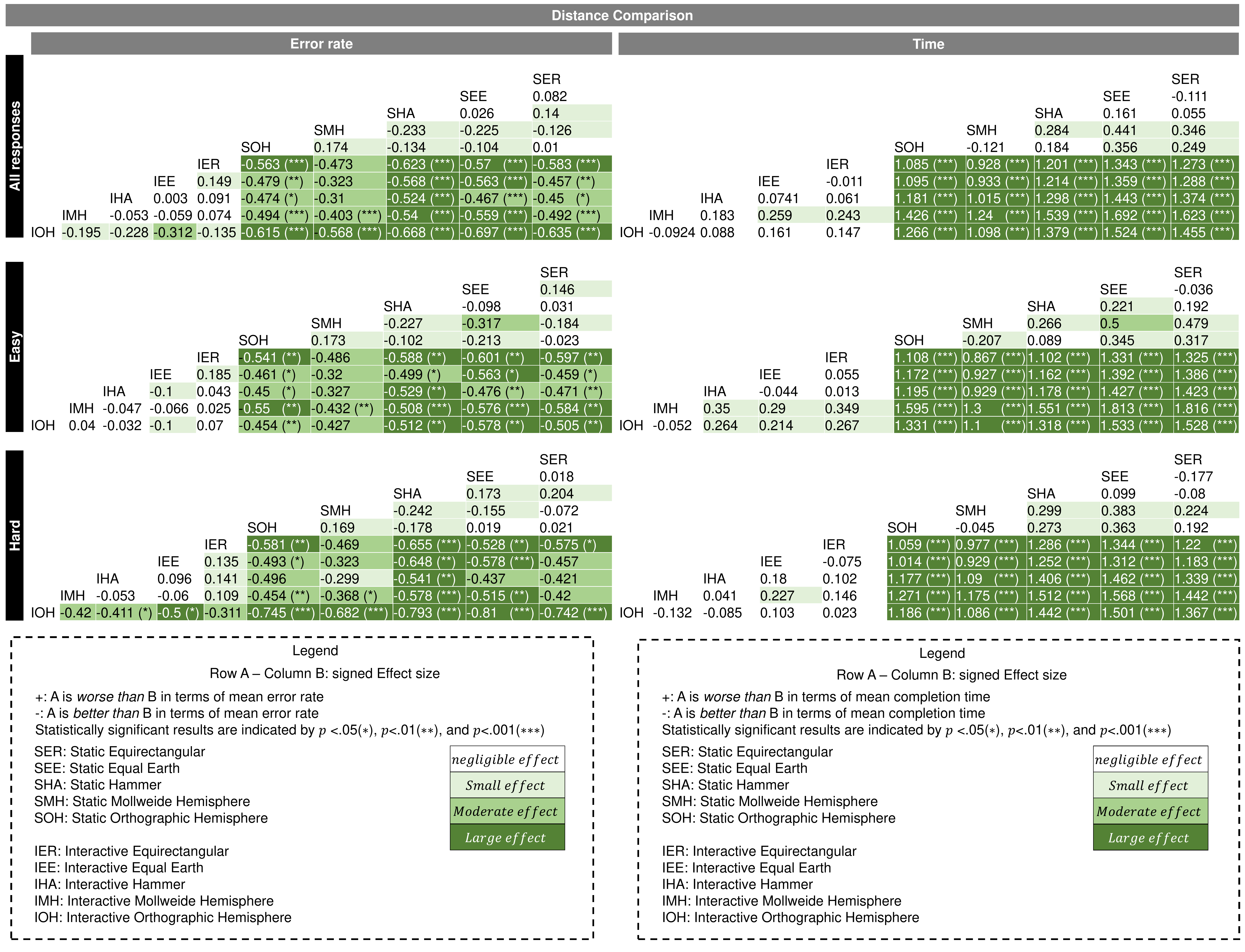}
    \Description[]{}
    \caption{Matrices of pairwise effect size results of \merror{} and \mtime{} of study 1 for Distance task. Values indicate effect size results of Cohen's r and Cohen's d~\cite{cohen2013statistical} for \merror{} and \mtime{}, respectively.}
    \label{fig:study-1-effect-size-distance}
\end{figure*}

\begin{figure*}
   \centering
    \includegraphics[width=1\textwidth]{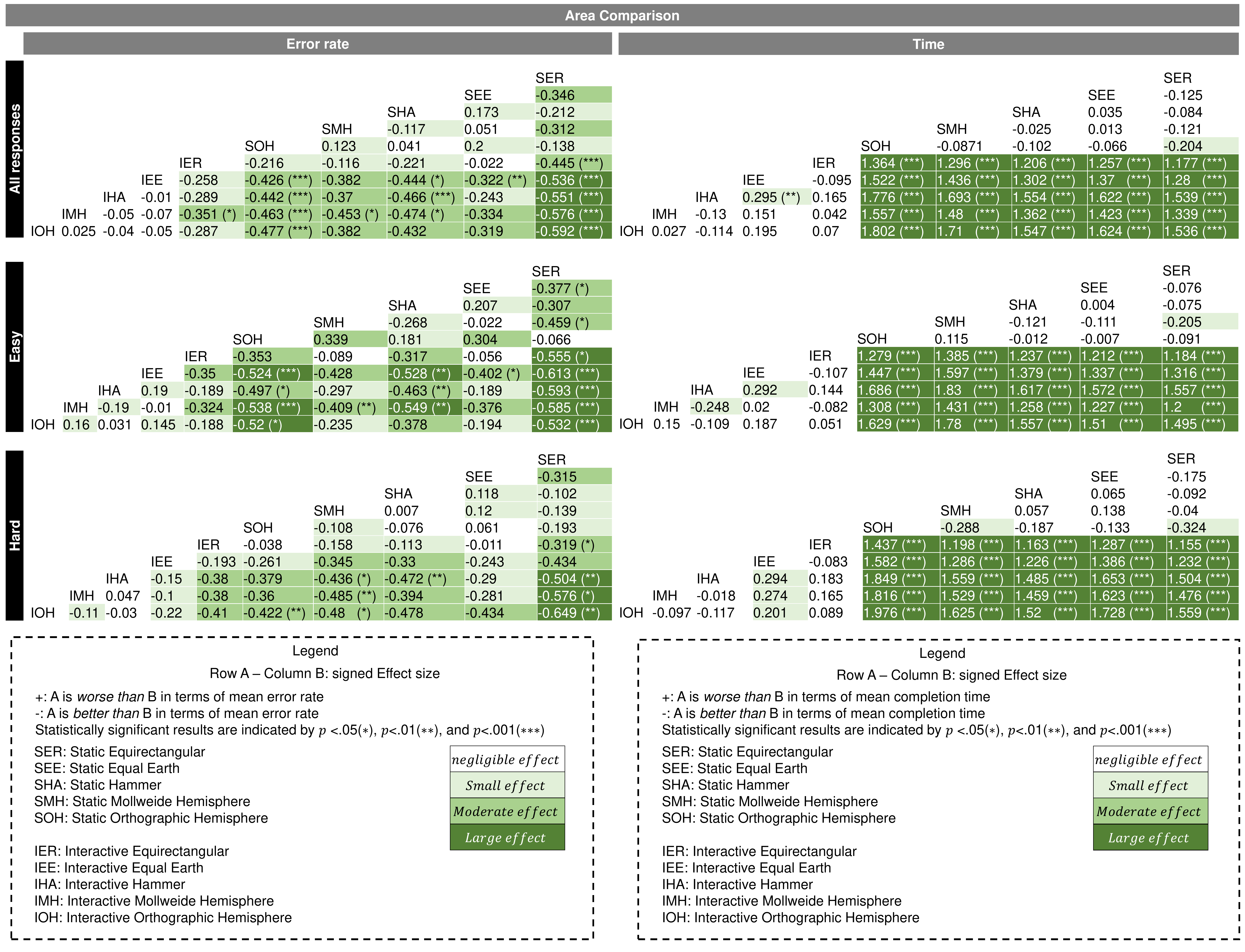}
    \Description[]{}
    \caption{Matrices of pairwise effect size results of \merror{} and \mtime{} of study 1 for Area task. Values indicate effect size results of Cohen's r and Cohen's d~\cite{cohen2013statistical} for \merror{} and \mtime{}, respectively.}
    \label{fig:study-1-effect-size-area}
\end{figure*}

\begin{figure*}
   \centering
    \includegraphics[width=1\textwidth]{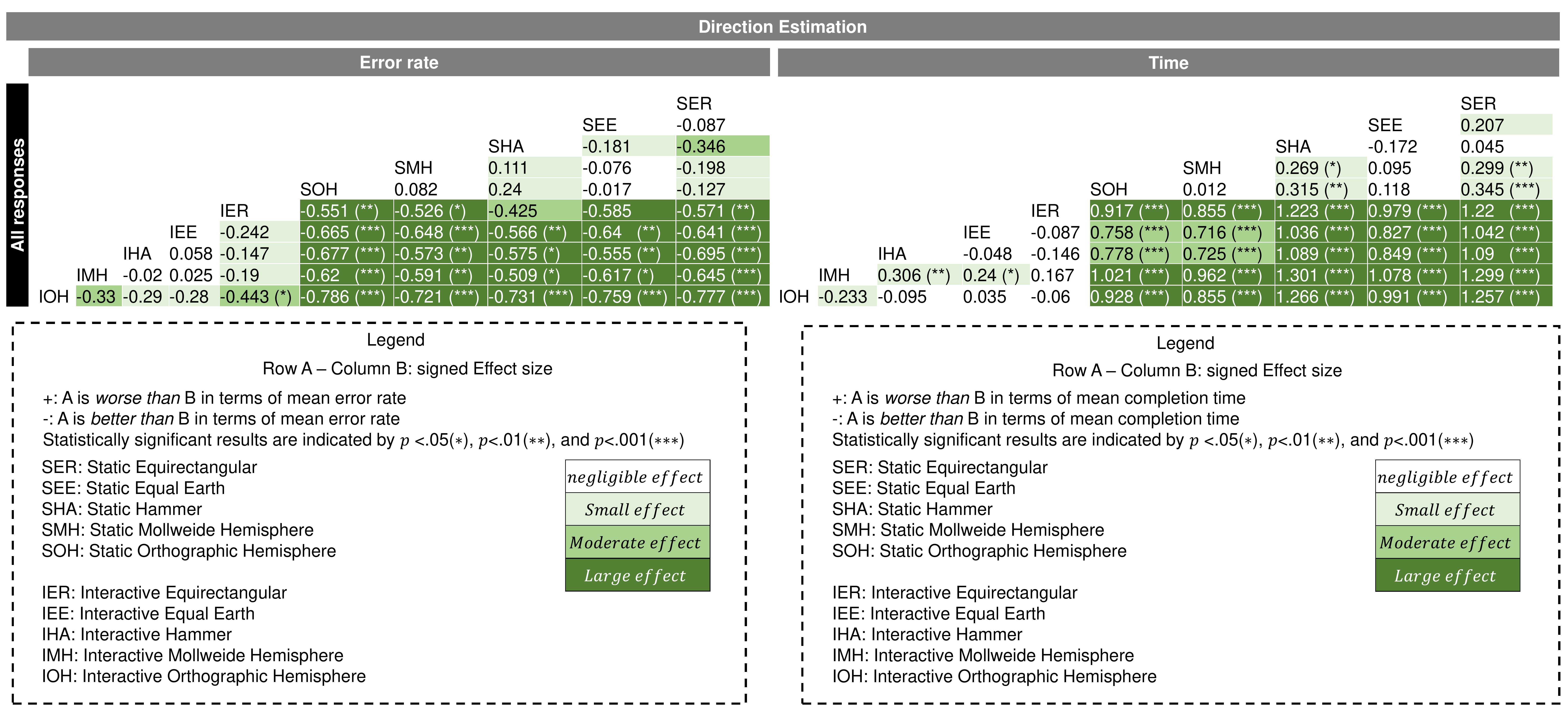}
    \Description[]{}
    \caption{Matrices of pairwise effect size results of \merror{} and \mtime{} of study 1 for Direction task. Values indicate effect size results of Cohen's r and Cohen's d~\cite{cohen2013statistical} for \merror{} and \mtime{}, respectively.}
    \label{fig:study-1-effect-size-direction}
\end{figure*}

\begin{figure*}
    \centering
    \includegraphics[width=1\textwidth]{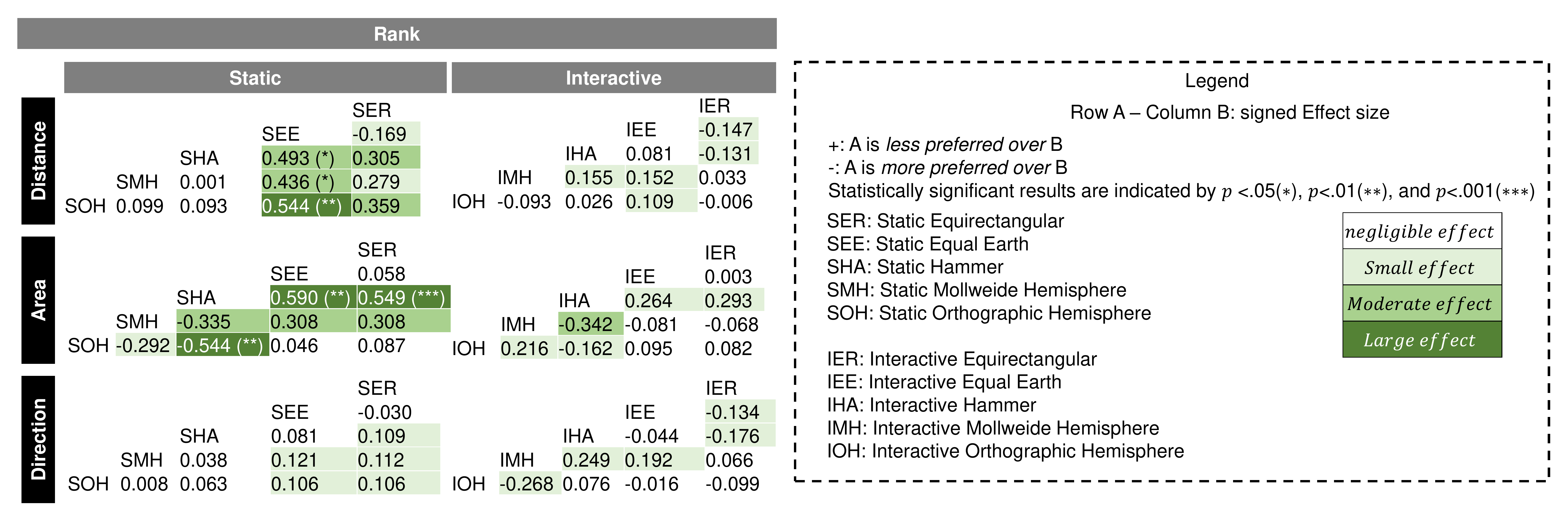}
    \Description[Legend]{}
    \caption{Matrices of pairwise effect size results of subjective user rank of Study 1 map projections within the \emph{static} (left) and \emph{interactive} (right) groups for three tested tasks. Values indicate effect sizes results of Cohen's r.}
    \label{fig:study-1-ranking-results-effect-size}
\end{figure*}

\subsection{Study 2: Network Projection Readability}
\label{sec:supplestudy2}

\begin{table*}
  \centering
  \begin{tabularx}{\linewidth}{X r r r r r r}
    \toprule
    {\small\textit{Projections}}
    &
    {\small\textit{\vtop{\hbox{Number of Graphs}}}}
    &
    {\small\textit{Datasets}}
    &
    \multicolumn{2}{c}{\textit{Number of Pixels}}
    &
    \multicolumn{2}{c}{\textit{\vtop{\hbox{Crossing Counts}}}}
    \\
    \midrule
    {\small\textit{}}
    & {\small \textit{\vtop{\hbox{}}}}
    &
    {\small\textit{}}
    & {\small \textit{\vtop{\hbox{No Pan}}}}
    & {\small \textit{\vtop{\hbox{Auto-Pan}}}}
    & {\small \textit{\vtop{\hbox{No Pan}}}}
    & {\small \textit{\vtop{\hbox{Auto-Pan}}}}
    
    \\
    \tequalearth{} & 10 & \dsmalleasy{}, \dsmallhard{} & 20916.3 &  23464.24 & N/A   & N/A  \\
    \hline
    \torthographic{} & 10 & \dsmalleasy{}, \dsmallhard{} & N/A &  N/A  & 262.16   & 208.7  \\
    \bottomrule
  \end{tabularx}
  \Description[Automatic panning results]{Automatic panning reduces mean number of pixels at the boundary, number of edges crossing the boundaries compared with no panning for \dsmall{} graphs at high (0.4) and low (0.3) modularity (with layout pre-computed by Algorithm 1);}
  \caption{Automatic panning results: mean number of pixels at the boundaries (higher values indicate less wrapping), and number of edges crossing the boundaries of 10 random runs of study graphs for cluster understanding tasks, with layouts pre-computed by algorithms in Sect.~\ref{sec:layoutalgorithms}.}~\label{tab:autopanresults}
\end{table*}

For \tclusteridentification{}, we found Layout had a significant effect on error rate in All Responses ($***$), Easy ($***$), Hard ($***$), Small ($***$), and Large ($***$). \ttorus{} ($***$), \tequalearth{} ($***$) and \torthographic{} ($*$) were more accurate than \tnodelink{}. \ttorus{} ($*$) and \tequalearth{} ($**$) were more accurate than \torthographic{}. We also found \ttorus{} and \tequalearth{} were more accurate than \tnodelink{} for Easy ($***$) and \torthographic{} for Easy ($***$).  We found \ttorus{}, \tequalearth{}, and \torthographic{} were more accurate than \tnodelink{} for Hard ($***$), Small ($***$). We also found \ttorus{} ($***$), \tequalearth{} ($***$) and \torthographic{} ($*$) were more accurate than \tnodelink{} for Large. \ttorus{} ($*$) and \tequalearth{} ($**$) were more accurate than \torthographic{} for Large.

We found Layout ($***$), Layout $\times$ Difficulty ($***$) and Layout $\times$ Size ($*$) had a significant effect on time. In Layout, \tnodelink{} is faster than \ttorus{}, \tequalearth{} and \torthographic{} (Layout $***$). \ttorus{} was found faster than \torthographic{} (Layout $***$). \tequalearth{} was faster than \ttorus{} and \torthographic{} (Layout $***$). In Layout $\times$ Difficulty, \tnodelink{} is faster than \ttorus{} (Easy/Hard $***$), \tequalearth{} (Easy $*$, Hard $***$), and \torthographic{} (Easy/Hard $***$). \tequalearth{} is faster than Torus (Easy $***$, Hard $*$) and \torthographic{} (Easy/Hard $***$). We also found \ttorus{} was faster than \torthographic{} (Easy $*$). In Layout $\times$ Size, \tnodelink{} was faster than \ttorus{} (Small/Large $***$), \tequalearth{} (Large $***$), and \torthographic{} (Small/Large $***$). \tequalearth{} was found faster than \ttorus{} (Small $**$, Large $***$), and \torthographic{} (Small/Large $***$). \ttorus{} was also found faster than \torthographic{} (Small $***$).

For user ranking, participants preferred \ttorus{} to \tnodelink{} ($**$) and \torthographic{} ($***$). Participants preferred \tequalearth{} to \tnodelink{} ($*$) and \torthographic{} ($***$).

For \tshortestpath{}, we found Layout had a significant effect on error rate in all ($***$). We found for All Responses, \tnodelink{} was more accurate than \tequalearth{} ($**$) and \torthographic{} ($***$). \ttorus{} was more accurate than \tequalearth{} ($**$) and \torthographic{} ($**$). In Easy, we found \tnodelink{} was more accurate than \tequalearth{} ($***$) for Easy, and \torthographic{} ($***$) for Easy. \ttorus{} was also more accurate than \tequalearth{} ($***$) for Easy and \torthographic{} ($**$) for Easy. We did not find significant difference in Hard.

We found Layout had a significant effect on time ($***$). \tnodelink{} was faster than \ttorus{}, \tequalearth{} and \torthographic{} ($***$). In Hard, we also found \tnodelink{} was faster than \ttorus{}, \tequalearth{} and \torthographic{} ($***$). We did not find significant difference in Easy.
For user ranking, participants preferred \tnodelink{} to \tequalearth{} ($**$) and \torthographic{} ($**$). \ttorus{} was also found preferred to \tequalearth{} ($***$) and \torthographic{} ($***$).

\subsubsection{Qualitative Feedback} 
\label{sec:supple_study2_qualitativefeedback}

\textbf{For \tclusteridentification{}, \tequalearth{} and \ttorus{} performed equally well and \tequalearth{} was slightly faster than \ttorus{}, while they both significantly outperformed \torthographic{} in terms of \merror{}, \mtime{}, and \mpref{}}.
\tequalearth{} and \ttorus{} were found to be easier to make out cluster boundaries, while the separated maps in \torthographic{} made it confusing, as participants mentioned \emph{``It was easier seeing the boundaries of the set number of clusters in [torus] representation.''} (P11), and \emph{``I found understanding the movement of the orthographic diagram quite challenging.''} (P4). 

Some participants explicitly mentioned that interaction improves the readability of clusters in \ttorus{} and \tequalearth{}, while \torthographic{} is confusing although it looks naturalistic, e.g., \emph{``Torus is the easiest to interpret because the clusters do not get distorted if the map is moved around. Equal Earth is actually easier than it looks to interpret because even though the clusters get distorted, if they are moved around, they actually bunch up together once they are put near the middle of the map. Orthographic Hemisphere uses 2 maps so it is much more difficult to interpret than the others due to this making it harder to isolate clusters.''} (P12), \emph{``Second place is Torus because you can very easily navigate. Third place is equal earth because you can see almost all of the clusters at the same time, but it's a bit confusing because of distortion. Last place is Ortographic Hemisphere because it is very confusing''} (P16), and \emph{``Orthographic was the most convenient one due to familiar shape of earth.''} (P6). 

\textbf{\tequalearth{}, \torthographic{} and \ttorus{} significantly resulted in less error than \tnodelink{}. \tequalearth{} and \ttorus{} were significantly preferred over \tnodelink{} but \ttorus{} and \torthographic{} took longer time than \tnodelink{}.}
Some participants explicitly mentioned that good separation of clusters in continuous surfaces such as \ttorus{} and \tequalearth{} helped understanding, e.g., \emph{``Distance between the clusters as well as perspective played a huge role in my choices. The more further apart the clusters were the easier it was for me to count them.''} (P18), \emph{``I think "equal earth" was the easiest and most understandable."Node-link" was something like a challenge for me,it was very difficult to understand it.''} (P28), \emph{``I could see patterns more easily in the torus representation. It was easier seeing the boundaries of the set number of clusters in that representation.''} (P11), and \emph{``Torus seems to be the easiest, because it was just flat surface with possibility of dragging.''} (P14), while  \tnodelink{} tended to appear tangled, e.g., \emph{``With the [\tnodelink{}], it was sometimes apparent where the clusters were, but I feel like I chose 1 cluster as an answer too frequently with the tightly packed examples.''} (P11).

\textbf{For \tshortestpath{}, \tnodelink{} and \ttorus{} performed equally well, both significantly outperforming \tequalearth{} and \torthographic{} in terms of \merror{} and \mpref{}, while \tnodelink{} is faster than all representations.}
Although \ttorus{} involves broken links across the boundaries, it appears similar to \tnodelink{} using straight links while the distortion of paths in \tequalearth{} might hamper path tracing tasks: \emph{``I am more familiar with the [\tnodelink{}] and torus shaped representations.''} (P6), and \emph{``The flatter that it seemed, the easier it was.''} (P20).

On the other hand, \torthographic{} has less link distortion but the interruption between two hemispheres and the strong perspective distortion near the boundary of the maps may make participants confused. For example, \emph{``Orthographic Hemisphere was even trickier because I couldn't get some of the dots on the same sphere sometimes and the time ran out.'' (P12)} and \emph{``Orthographic Hemisphere is the last because I found it the most restrictive, since it is difficult to make the red points appear together since the curves of the links don't bend, making it harder to see all the ways they connect.''} (P21).

Although auto-pan provides some benefits for cluster separation for \tclusteridentification{}, it seems it has less benefits for \tshortestpath{}, as participants mentioned they still need to position the image to see the full path to identify the shortest path for connecting the nodes within the time limit for \tshortestpath{}. For example, \emph{``Equal earth and orthographic hemisphere would make things more puzzling for me. When I had to wrap the map in order to find the fastest path for the nodes, it would usually take me more time to estimate and find the lines between the two red nodes. I would say that when curves are being shaped it is more complex to find the fastest path. It also made me anxious because I had to answer nearly at the last moment. Torus was almost as the simple node-link diagram. However, I found it easier to find the links between the nodes in a diagram that I had little or no interaction.''} (P35) and \emph{``I was more confident using the Node-Link. It was simpler to use visually. With the other three, there were times when it was difficult to position the map in a way that I could identify the shortest path for connecting the nodes.''} (P45).

\begin{figure*}
    \centering
    \includegraphics[width=1\linewidth]{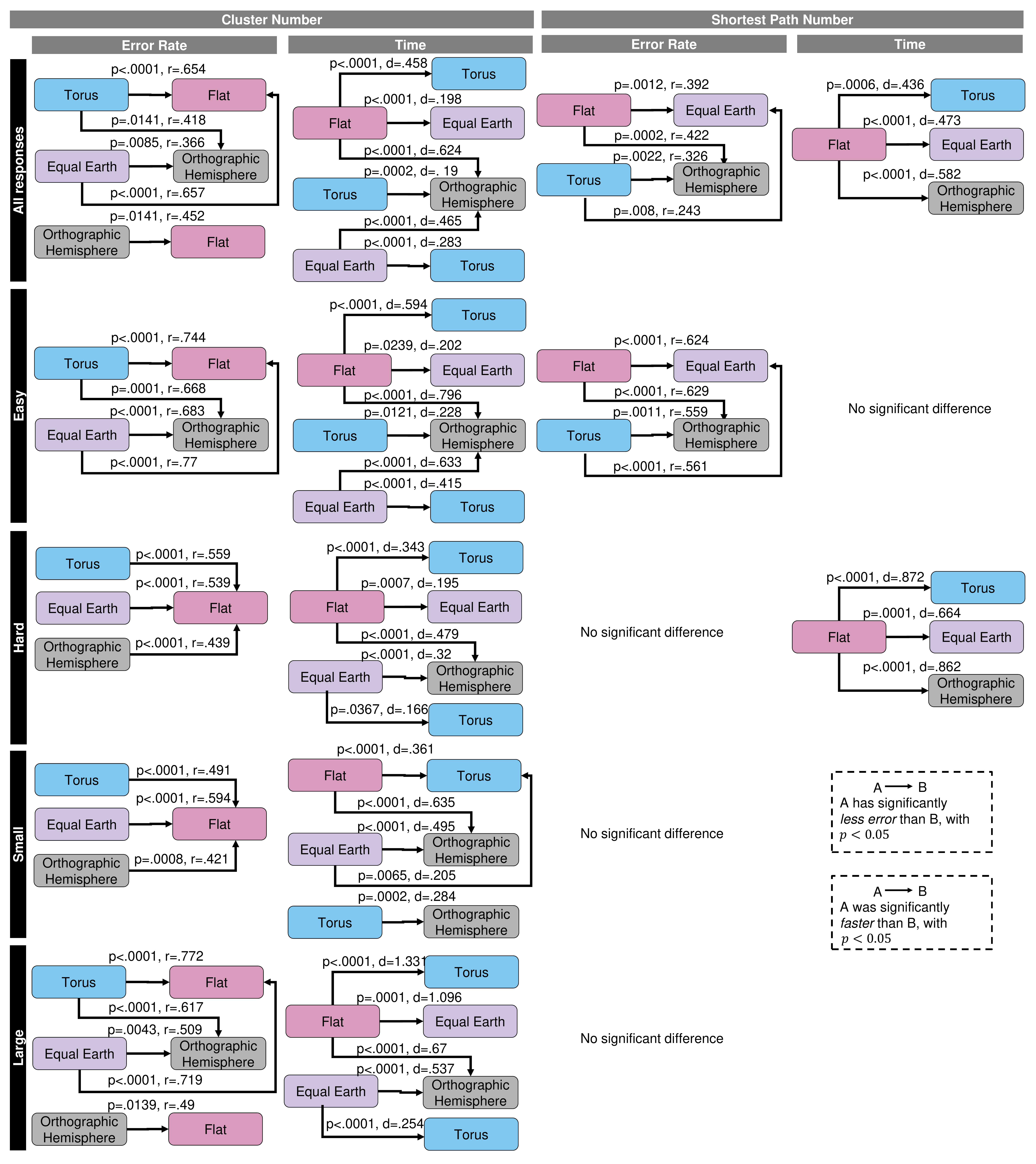}
    \Description[Legend]{}
    \caption{Study 2: Statistically significant results of \merror{} and \mtime{} of study 2 for Cluster Number task (left) and Shortest Path Number task (right). Effect size results of Cohen’s r and Cohen’s d for \merror{} and \mtime{} are presented along with the arrows.}
    \label{fig:study-2-results-graphics}
\end{figure*}

\begin{figure*}
    \centering
    \includegraphics[width=1\linewidth]{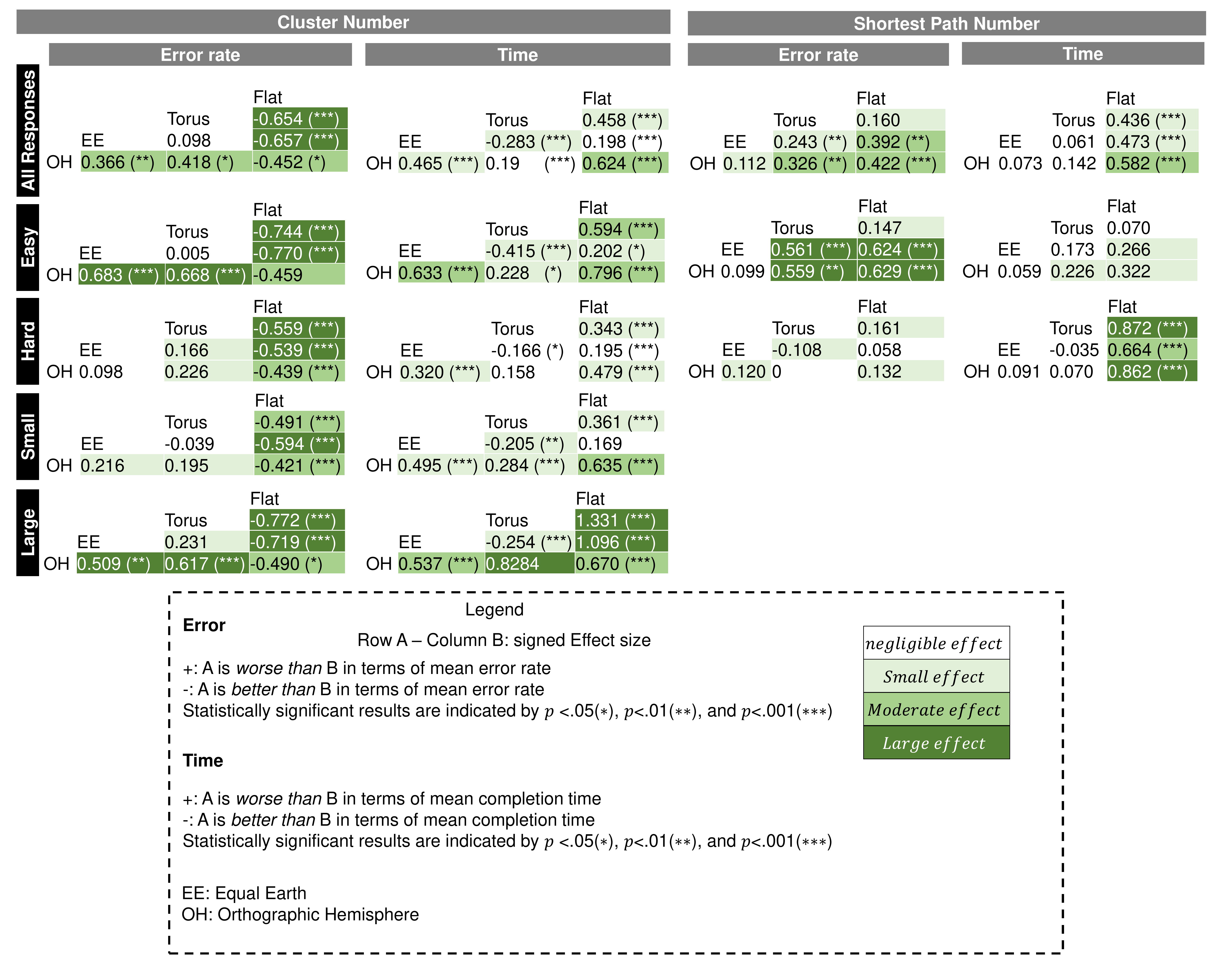}
    \Description[Legend]{}
    \caption{Matrices of pairwise effect size results of \merror{} and \mtime{} of study 2 for Cluster Number task (left) and Shortest Path Number task (right). Values indicate effect size results of Cohen's r and Cohen's d~\cite{cohen2013statistical} for \merror{} and \mtime{}, respectively.}
    \label{fig:study-2-results-effect-size}
\end{figure*}

\begin{figure*}
    \centering
    \includegraphics[width=1\linewidth]{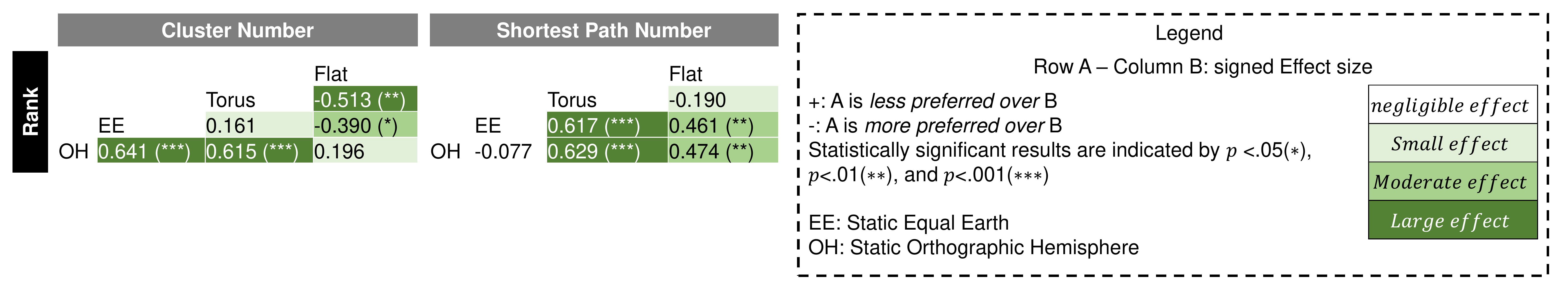}
    \Description[Legend]{}
    \caption{Matrices of pairwise effect size results of \mpref{} of study 2 for Cluster Number task (left) and Shortest Path Number task (right). Values indicate effect size results of Cohen's r.}
    \label{fig:study-2-results-rank-effect-size}
\end{figure*}

\end{document}